\documentclass[usenatbib,useAMS]{mnras}

\usepackage{graphicx}	
\usepackage{amsmath}	
\usepackage{amssymb}	
\usepackage{multicol}        
\usepackage{bm}		
\usepackage{pdflscape}	
\usepackage{dblfloatfix}    
\usepackage[T1]{fontenc}
\usepackage{ae,aecompl}

\usepackage{newtxtext,newtxmath}

\title{SDSS-IV MaStar: Theoretical Atmospheric Parameters for the MaNGA Stellar Library}

\author[L. Hill]{Lewis Hill$^{1}$\thanks{E-mail:lewis.hill@port.ac.uk},
Daniel Thomas$^{1,2}$,
Claudia Maraston$^{1}$,
Renbin Yan$^{3}$,
Justus Neumann$^{1}$,
\newauthor
Andrew Lundgren$^{1}$,
Daniel Lazarz$^{3}$,
Yan-Ping Chen$^{4}$,
Michele Cappellari$^{5}$,
Jon A. Holtzman$^{6}$,
\newauthor
Julie Imig$^{6}$,
Katia Cunha$^{7,8}$,
Guy Stringfellow$^{9}$,
Dmitry Bizyaev$^{10,11}$,
David R. Law$^{12}$,
\newauthor
Keivan G. Stassun$^{13}$,
Niv Drory$^{14}$,
Michael Merrifield$^{15}$,
Timothy C. Beers$^{16}$
\\
$^{1}$Institute of Cosmology and Gravitation, University of Portsmouth, Burnaby Road, Portsmouth, PO1 3FX, UK
\\
$^{2}$ School of Mathematics and Physics, University of Portsmouth, Lion Gate Building, Portsmouth, PO1 3HF, UK
\\
$^{3}$Department of Physics and Astronomy, University of Kentucky, 505 Rose Street, Lexington, KY, 40506-0055, USA
\\
$^{4}$ New York University Abu Dhabi, Abu Dhabi, P.O. Box 129188, United Arab Emirates\\
$^{5}$ Sub-Department of Astrophysics, Department of Physics, University of Oxford, Denys Wilkinson Building, Keble Road, Oxford, OX1 3RH\\
$^{6}$ Department of Astronomy, New Mexico State University, Las Cruces, NM 88003, USA\\
$^{7}$ Observat\'{o}rio Nacional/MCTIC, R. Gen. Jos\'{e} Cristino, 77, 20921-400, Rio de Janeiro, Brazil\\
$^{8}$ Steward Observatory, University of Arizona, 933 North Cherry Avenue, Tucson, AZ 85721-0065, USA\\
$^{9}$ Center for Astrophysics and Space Astronomy, Department of Astrophysical and Planetary Sciences, University of Colorado, 389 UCB, Boulder, CO 80309-0389, USA
\\
$^{10}$ Apache Point Observatory and New Mexico State University, P.O. Box 59, Sunspot, NM 88349, USA\\
$^{11}$ Sternberg Astronomical Institute, Moscow State University, Universitetskij pr. 13, Moscow, Russia\\
$^{12}$ Space Telescope Science Institute, 3700 San Martin Dr., Baltimore, MD 21218, USA\\
$^{13}$ Vanderbilt University
Physics and Astronomy Dept. 
6301 Stevenson Center Ln.
Nashville, TN 37235, USA\\
$^{14}$ McDonald Observatory, The University of Texas at Austin, 1 University Station, Austin, TX 78712, USA\\
$^{15}$ School of Physics and Astronomy, University of Nottingham, University Park, Nottingham, NG7 2RD, UK\\
$^{16}$ Department of Physics and JINA Center for the Evolution of the Elements, University of Notre Dame, Notre Dame, IN 46556, USA
}

\date{Accepted XXX. Received YYY; in original form ZZZ}

\pubyear{2020}

\begin{document}
\label{firstpage}
\pagerange{\pageref{firstpage}--\pageref{lastpage}}
\maketitle

\begin{abstract}
We calculate the fundamental stellar parameters effective temperature, surface gravity and iron abundance - T$_{\rm eff}$, log g, [Fe/H] - for the final release of the Mapping Nearby Galaxies at APO (MaNGA) Stellar Library (MaStar), containing 59,266 per-visit-spectra for 24,290 unique stars at intermediate resolution ($R\sim1800$) and high S/N (median = 96). We fit theoretical spectra from model atmospheres by both MARCS and BOSZ-ATLAS9 to the observed MaStar spectra, using the full spectral fitting code pPXF. We further employ a Bayesian approach, using a Markov Chain Monte Carlo (MCMC) technique to map the parameter space and obtain uncertainties. Originally in this paper, we cross match MaStar observations with Gaia photometry, which enable us to set reliable priors and identify outliers according to stellar evolution. In parallel to the parameter determination, we calculate corresponding stellar population models to test the reliability of the parameters for each stellar evolutionary phase. We further assess our procedure by determining parameters for standard stars such as the Sun and Vega and by comparing our parameters with those determined in the literature from high-resolution spectroscopy (APOGEE and SEGUE) and from lower-resolution matching template (LAMOST). The comparisons, considering the different methodologies and S/N of the literature surveys, are favourable in all cases. Our final parameter catalogue for MaStar cover the following ranges: $2592 \leq $ T$_{\rm eff} \leq 32983\;$K; $-0.7 \leq $ log g $ \leq 5.4\;$dex; $-2.9 \leq $ [Fe/H] $\leq 1.0\;$dex and will be available with the last SDSS-IV Data Release, in December 2021.
\end{abstract}




\begingroup
\let\clearpage\relax

\endgroup
\newpage

\section{Introduction}
The age and chemical distribution of unresolved stellar populations in galaxies and star clusters can be probed via evolutionary population synthesis (EPS) modelling \citep{tinsley1972_galevo, tinsley1980, bruzual83, renzini86, maraston98, bruzual_charlot03, maraston05, leitherer99, vazdekis96, vazdekis10, vazdekis12, fioc97, conroy09, maraston_stromback11, maraston20, thomas03, thomas11}. This technique relies on stellar evolution theory to model the spectral energy distribution (SED) of stellar systems across the Universe. The basis of EPS models is the Simple Stellar Population (SSP), which is a collection of stars that are coeval and chemically homogeneous. To create a model SSP there are three fundamental inputs: the stellar initial mass function (IMF), stellar evolutionary tracks and a library of stellar spectra with known atmospheric parameters and a comprehensive coverage of the different stages of stellar evolution.

With respect to the spectral library input, one has a selection of empirical and theoretical libraries to choose from. Empirical libraries include: ELODIE \citep{prugniel01}, MILES \citep{miles06}, X-SHOOTER Spectral Library \citep[XSL,][]{chen14, gonneau20}, STELIB \citep{stelib03}, PICKLES \citep{pickles98} and GRANADA \citep{martins05}. These libraries differ vastly in their coverage of stellar parameters, in their spectral resolution, and wavelength range. Not all of them are adequate to model the SEDs of galaxies from modern surveys such as Mapping Nearby Galaxies at Apache Point Observatory (MaNGA) \citep{manga15, yan16a, yan16b, blanton17, drory15, law15, law16}, which has collected IFU spectroscopy for over 10,000 nearby galaxies (z $\sim0.05$). In most cases the wavelength coverage is not sufficient to model the entire SED wavelength range. In response to this gap in the market, the MaNGA Stellar Library (MaStar, \citep{yan19}) has been established, covering a larger parameter space in atmospheric properties, mass and luminosity than any other spectral library. Such comprehensive coverage will allow for the creation of robust SSP models. Furthermore, MaStar collects spectra using the Baryon Oscillation Spectroscopic Survey (BOSS) spectrographs \citep{smee13} in a wavelength range of $3620 - 10350\;$\AA\ and median resolution of $R\sim1800$. This means that the SSP models created with this library will be perfectly suited for Sloan Digital Sky Survey  (SDSS, \citep{york2000, eisenstein2011, blanton17}) spectra and spectroscopic surveys also at high redshift.

To calculate EPS models using empirical spectra, one needs accurate stellar atmospheric parameters for the empirical spectra to be able to link them to the theoretical parameters expected for stellar evolutionary phases. These parameters include the effective temperature T$_{\text{eff}}$, surface gravity log g and iron abundance [Fe/H] - hereafter the fundamental stellar parameters (FSPs). These FSPs can be determined by analysing the photometric and spectroscopic data of a star with theoretical spectra from atmosphere models.

\begin{figure}
 \includegraphics[width=0.95\columnwidth]{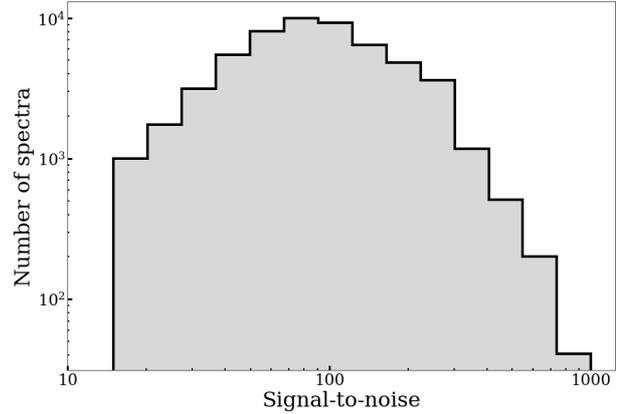}
 \caption{Histogram of the median signal-to-noise ratio per pixel of all MaStar good visit spectra in the MaNGA Product Launch 9 data release. The median value of this distribution is 96 (per pixel).}
 \label{fig:mastar_sn}
\end{figure}

Stellar parameter determination is not only essential for creating population models, but also for Milky Way modelling and our understanding of stellar astrophysics. Consequently, there have been many studies and algorithms devoted to the determination of the FSPs. For large-scale surveys, including MaStar, it has become necessary to automate this process, which is now possible thanks to the advancement in computer processing power. Existing work dedicated to this problem includes the APOGEE Stellar Parameter and Chemical Abundances Pipeline (ASPCAP, \citet{aspcap15, apogee}) which uses a $\chi^2$ minimisation approach in the software package \textit{FERRE}. They determine FSPs by fitting the observed spectra to synthetic spectra over the range of $\sim200\;$nm and chemical abundances are found by fitting in a narrow wavelength range around interpolated spectra. This analysis is made at a higher spectral resolution than MaStar at R $\sim22,500$. \textit{The Payne} \citep{ting19}, uses a fully connected neural network also trained on synthetic spectra. This approach relies on mapping up to 25 stellar labels to each spectrum and uses a least squares minimisation to train the network weights. Similar to this, but trained on empirical stellar spectra, is \textit{The Cannon} \citep{ness15}. This approach uses a transfer learning method of training a data-driven model on some empirical spectra with high-fidelity labels and applying the model to a custom library of survey spectra, matching the line spread function (LSF) to observations. Furthermore, the Universit\'e de Lyon Spectroscopic Analysis Software (ULySS, \citet{Koleva09}), which is based on an early IDL version of the penalized pixel-fitting method (pPXF, \citet{cappellari04, ppxf17}) takes the approach of $\chi^2$ minimisation, with empirical data provided by the ELODIE interpolator as their reference set and LSF matched after analysis. They first use convergence maps to identify which parameter combinations converge to the absolute minimum  $\chi^2$ and combinations which may lead to local minima. They then use Monte-Carlo simulations with $\chi^2$ maps to break any degeneracies and determine atmospheric parameters. Despite their success and accuracy, such methods cannot be used to predict the full parameter range of MaStar objects due to limitation in their model parameter grid and narrow wavelength ranges which are not suitable for the wide range of stellar spectral types presented in MaStar. Furthermore, by calculating parameters from fitting a wide wavelength range, we are less dependent on possible inaccuracies of individual lines in the model atmospheres.

\defcitealias{maraston20}{M20}

In this paper we present a method for the task of stellar parameter determination with which we derived the FSPs of the final catalogue of MaStar spectra (MPL11), including 59,266 per-visit spectra for 24,290 unique stars\footnote{Note that the full catalogue of parameters will be available at the official data release. To clarify for readers outside of the SDSS-MaNGA collaboration, the data are made available to the public through scheduled data releases (DRs) and circulated internally in a MaNGA Product Launch (MPL). Please refer to Table 1 of \citet{law21} for further details and the dates of each data release.}. This is an extension of the method we developed for the early MaStar data release\footnote{MaNGA Product Launch 7, hereafter MPL7, of 8,646 per-visit spectra for 3,321 stars as part of the SDSS Data Release 15, \citet{yan19, DR15}}, in support of the first MaStar-based stellar population models \citet[][ hereafter M20]{maraston20}. In both cases, we fit the observed spectra with a wide grid of theoretical spectra from model atmospheres using the pPXF full spectral fitting code \citep{cappellari04, ppxf17}. In this paper, we augment the template fitting with a Markov Chain Monte Carlo (MCMC) technique in order to thoroughly map the parameter space and derive uncertainties. For completeness, in Appendix \ref{sec:method1} we describe our previous method which uses a discrete $\chi^{2}$ approach and compare the results with \citet{chen2020}. In Appendix \ref{sec:comparison_methods} the discrete $\chi^{2}$ and MCMC methods are compared. The conclusion is that - in regard to stellar population performance - the two approaches are equivalently good.

It should be also noted that other efforts exist within MaStar to derive stellar parameters based on different techniques and methods \citep{chen2020}, Chen et al. (in preparation), Lazarz et al. (in preparation) and Imig et al. (in preparation). All parameters will be released via the SDSS value added catalogue alongside a median set in December 2021 and their comparison described in Yan et al. 2021, {\it in prep.}.

The structure of this paper is as follows. Section \ref{sec:data} briefly describes the main features and target selection of MaStar. This section also gives a description of the adopted theoretical stellar atmospheres and the pre-processing steps required for our analysis. In Section \ref{sec:method2} we describe our methodology and results. In Section \ref{sec:comparison_testing} we present stellar fits of the Sun and Vega to test our methodology against stars with well determined FSPs. In this section we also compare a subset of our parameters to other stellar parameter catalogues. In Section \ref{sec:ssp} we explain the link between this work and the population model calculations. Finally, the work is summarised and future considerations are made in Section \ref{sec:conclusion}.

\section{Data}
\label{sec:data}
\subsection{Observations}
\label{sec:observations}
The aim of MaStar is to collect stellar spectra covering a wider parameter space in T$_{\text{eff}}$, log g and [Fe/H] compared to previous libraries. Since this release MaStar now contains 59,266 high quality per-visit spectra for 24,290 stars (MPL11). The first public data release (DR15) is drawn from MPL7 of MaStar and contains 8,646 per-visit spectra for 3,321 stars \citep{yan19}. 

\begin{figure}
 \includegraphics[width=0.95\columnwidth]{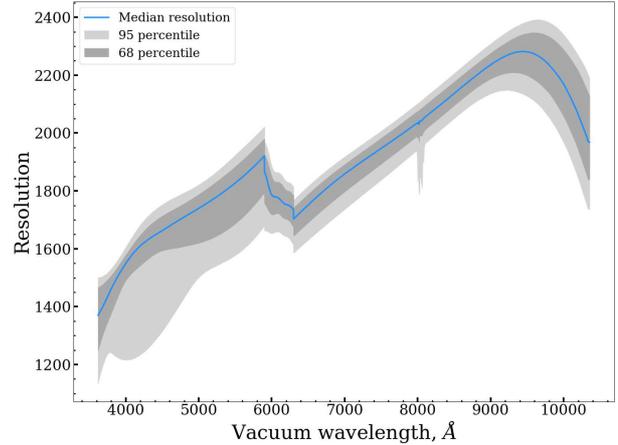}
 \caption{Median spectral resolution ($R = \lambda / \Delta \lambda$) of MaStar spectra (MPL11) as a function of wavelength (\AA). We also show the 68\% (dark grey) and 95\% (light grey) percentiles in resolution around the median. For further details of the features in this figure, please see \citet{law21}.}
 \label{fig:mast_res}
\end{figure}

Observations are carried out on the 2.5 m Sloan Foundation Telescope \citep{gunn06} located at the Apache Point Observatory. The MaStar library has been created thanks to parallel observations with the APOGEE-2N \citep{majewski16} survey by using MaNGA fibre bundles to take optical spectra in the same field of view. The use of fibre bundles allows for more accurate flux calibration compared to single fibre methods. Furthermore, MaStar can recover a higher signal to noise per object (median = 96 (per spectral pixel) in MPL11, see Figure \ref{fig:mastar_sn}) and wider wavelength coverage ($3620 - 10350$ \AA) compared to previous stellar spectroscopic surveys. The resolution of each individual spectrum varies due to changes in temperature of the fibers and because the focal plane and CCDs are not flat. In Figure \ref{fig:mast_res} we show the median, wavelength-dependent resolution for all MPL11 spectra, of which the median resolution ($R = \lambda / \Delta \lambda$) is approximately 1800. We also show the 1 and 2$\sigma$ range of resolution values about the median, indicated by the dark and light grey stripes. 
Details of the features in the resolution vector can be found in \citet{yan19} and \citet{law21}. Using the MaNGA fiber bundles and BOSS spectrographs means that stellar population models created with this library will be perfectly suited for the analysis of MaNGA, other SDSS spectra and a wide range of current and forthcoming spectroscopic surveys.

As mentioned, one of the main aims of the survey is to cover a parameter space in stellar properties (T$_{\text{eff}}$, log g and [Fe/H]) larger than previous efforts. This will contribute to creating more robust SSP models compared to previous empirical libraries. This has been achieved by firstly creating a photometry and astometry system based on the Pan-STARRS1 \citep{chambers16} and American Association of Variable Star Observers (AAVSO) Photometric All-Sky Survey (APASS\footnote{\url{https://www.aavso.org/apass/}}). Stars with known stellar parameters were identified in this catalogue and the coordinates used for targeting. In order to recover a uniform coverage in the FSPs and [$\alpha$/Fe], existing stellar parameter catalogues such as the Apache Point Observatory Galactic Evolution Experiment (APOGEE, \citet{aspcap15}), the Sloan Extension for Galactic Exploration and Understanding (SEGUE, \citet{segue1}) and the Large Sky Area Multi-Object Fiber Spectroscopic
Telescope (LAMOST, \citet{boeche18}) were then used to guide observations. This sophisticated target selection also ensures that the very hot and cold stars are observed as well as an oversampling of rare combinations of stellar parameters. The survey also observes many of the targets multiple times, this helps to account for any variability in the stellar atmosphere. Please refer to Section~3 of \citet{yan19} for more detail on MaStar target selection.

\subsection{Synthetic Stellar Atmospheres}
Most stellar parameter pipelines rely on comparing observations to ground truth data which has been parameterised with some level of confidence. For this purpose, we use the model atmosphere grids from MARCS \citep{gustafson08} and BOSZ-ATLAS9 \citep{meszaros12, bohlin17}. In general, theoretical stellar atmospheres are created by combining known atomic and molecular transitions with certain assumptions regarding properties such as local thermodynamic equilibrium (LTE), microturbulence ($\xi$) and model geometry. Synthetic spectra are then produced and compared to empirical observations.

By using synthetic spectra, we can ensure the full wavelength range of MaStar spectra can be modelled and resolution matched by downgrading to the MaStar wavelength-dependent resolution. In future iterations of our pipeline we plan to include models that account for the varying LSF in MaStar spectra. However, this will significantly increase the computation time. 

Using the MARCS model library we can cover the temperature range of $2500 - 8000\;$K, with log g values ranging from $-0.5$ to $5.5\;$dex and [Fe/H] from $-2.5$ to $1\;$dex. We use spherical model geometry between log g of $-0.5$ and $3$ and plane-parallel between log g of 3.5 and 5.5 for a more accurate description of the FSPs for giants and dwarfs. We also ensure the microturbulence parameter is set to $2$ kms$^{-1}$ for all MARCS models for homogeneity with the BOSZ-ATLAS9 assumptions (see below). The MARCS models also assume local thermodynamic equilibrium (LTE) and use the standard mixing-length model for convection. The synthetic spectra generated by these models were downloaded at a resolution of $R = 20,000$ and downgraded to the MaStar resolution.

The version of BOSZ-ATLAS9 model library covers temperatures of $3500 - 35,000\;$K, log g values range from $0$ to $5$ dex and [Fe/H] from $-2.5$ to $0.5\;$dex. These models are of a plane-parallel geometry with a microturbulence of $2$ kms$^{-1}$. The downloaded resolution is $R = 5,000$ before downgrading to the MaStar resolution.

\begin{figure}
 \includegraphics[width=0.9\columnwidth]{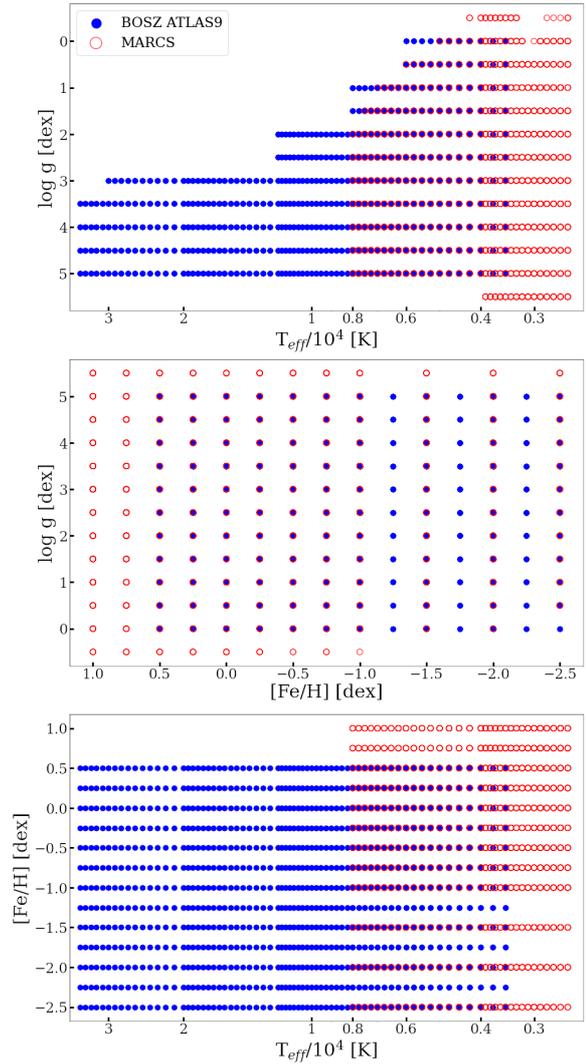}
 \caption{Adopted grids of MARCS (red) and BOSZ-ATLAS9 (blue).}
 \label{fig:marcs-atlas_grid}
\end{figure}

We allow some extrapolation of the models by extending up to $40,000\;$K in T$_{\rm eff}$, to $-1$ and $6\;$dex in log g and down to $-3\;$ dex for [Fe/H]. The extension to $40,000\;$K is justifiable as at these temperatures, the spectrum is close to that of a black-body. The extrapolation of log g and [Fe/H] will have a minor effect on the width and depth of absorption lines. This combination of parameters gives 3,907 and 5,891 MARCS and BOSZ-ATLAS9 spectra, respectively. We then separate the grids based on the flat priors described in (see Section \ref{sec:mcmc_priors}). If the minimum value of the T$_{\rm eff}$ prior is greater then $5000\;$K, then only BOSZ-ATLAS9 models are used. Otherwise for cooler spectra, both grids are explored separately and the best fitting model of the two grids is selected based on $\chi^2$. This is done to avoid the issue of interpolating between non-continuous synthetic spectra at the model grid interface. The grid of model parameters can be found in Figure \ref{fig:marcs-atlas_grid}.

In our analysis we focus on the main atmospheric parameters required for creating stellar population models: T$_{\rm eff}$, log g and [Fe/H]. However, there is scope to widen this parameter space to other properties such as stellar rotation, microturbulence or element abundance ratios, to name a few. Such properties will affect parameter determination due to relationships inherent in the spectra. In this first paper we use solar-scaled models at all metallicities, \textit{i.e.} no alpha-element enhancement so that all models are homogeneous and comparable regarding this parameter. We shall tackle the determination of detailed element abundances in future works. A constant value of $\xi = 2$ kms$^{-1}$ is adopted as this is what is available in the BOSZ models (as also used in previous works e.g. \citet{castelli94}). We also assume no rotational velocity.

To secure ourselves that these assumptions do not include strong systematics in our derived parameters, we assess the effect by $\xi$ and rotational velocity in Appendix \ref{sec:additional_params}. The result is that varying $\xi$ mainly leads to an offset in T$_{\rm eff}$ but the difference is generally small ($\pm100\;$K) and within our parameter errors. For rotation, we see no relationship between velocity and gravity. Accounting for these effects is necessary when deriving properties from individual lines at high resolution, but for the moderate resolution of MaStar and because we fit over a wide wavelength range, it seems negligible.

\subsection{Pre-processing}
\begin{figure}
 \includegraphics[width=0.95\columnwidth]{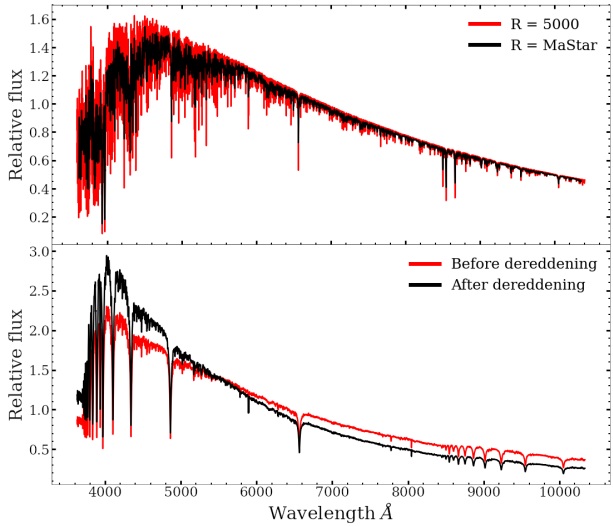}
 \caption{Top panel: An example BOSZ-ATLAS9 spectrum (T$_{\rm eff}$ = $5750\;$K, log g$ = 4$, [Fe/H] $= 0\;$dex) before and after downgrading to MaStar resolution. Bottom panel: An example MaStar spectrum before and after it has been de-reddened with $E(B-V) = 0.19$ using the \citet{Fitzpatrick99} dust extinction law. All spectra shown are median normalised after processing.}
 \label{fig:preprocess}
\end{figure}

Before MaStar and synthetic spectra can be compared, some preprocessing is required. We firstly downgrade the models of MARCS and BOSZ-ATLAS9 to the same resolution vector as the median resolution of MaStar data (see Figure \ref{fig:mast_res} for the MaStar resolution vector). As the resolution is wavelength-dependent, we cannot use a standard Gaussian kernel to downgrade the synthetic spectra. Instead, we use the \textit{'gaussian\_filter1d'} function provided in the pPXF package. This allows for a variable sigma (standard deviation of the Gaussian) for every pixel and is computationally efficient. The models are also resampled at the same velocity sampling as the data, which for the SDSS spectrographs is $69\;$km/s/pixel. This is done in order to make a pixel-by-pixel comparison later in the analysis. To resample, we use the python package \texttt{SpectRes} \citep{spectres17} which conserves flux density per angstrom. An example of original and downgraded model is in the upper panel of Figure \ref{fig:preprocess}.

\begin{figure}
 \includegraphics[width=0.95\columnwidth]{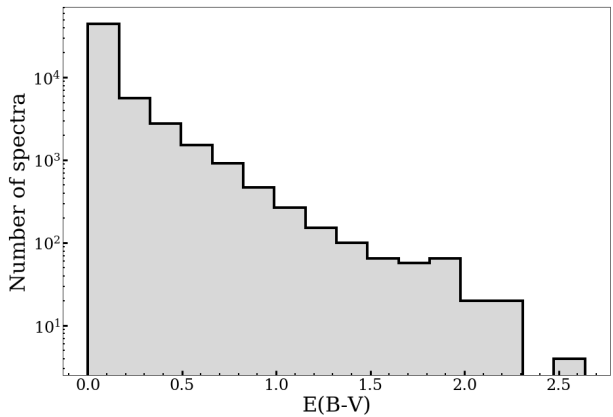}
 \caption{Distribution of $E(B-V)$ of MaStar data derived from the 3D dust maps of \citet{green19...93G}.}
 \label{fig:mast_ebv}
\end{figure}

Furthermore, the observed spectra require some correction for dust extinction, which is also referred to as being de-reddened. The MaStar catalogue is matched to the Gaia DR2 \citep{gaia18} parallax values, which for 24,290 unique stars results in 23,180 clean matches (\textit{i.e.} only $\sim5$ per cent of observations lack Gaia parallax values). The Gaia crossmatch also allows us to create priors based on their reported photometry, as explained in Section \ref{sec:mcmc_priors}. Distance estimates are then used from \citet{bailer-jones18} in combination with the 3D dust map provided by \citet{green19...93G}, who use data from the Gaia, Pan-STARRS 1 and 2MASS \citep{2mass06} surveys to obtain accurate values of $E(B-V)$. With the calculated $E(B-V)$ values, extinction is corrected for by using the \citet{Fitzpatrick99} dust extinction function for each observation. See the bottom panel of Figure \ref{fig:preprocess} which shows a MaStar spectrum being de-reddened according to an $E(B-V) = 0.19$. The distribution of $E(B-V)$ values for the target stars can be found in Figure \ref{fig:mast_ebv}. The spectra without a Gaia crossmatch or Bailor-Jones distance, hence without 3D dust map $E(B-V)$ values are not used in our final parameter catalogue.

\section{MCMC}
\label{sec:method2}
Here we describe the use of Markov Chain Monte Carlo (MCMC) methods in parallel with the full spectral fitting code pPXF for parameter determination of the MPL11 catalogue of stellar spectra. With this method, we are able to use $\chi^2$ statistics as an indication of the quality of fit. These values are used to clean the data by discarding parameters corresponding to bad fits with high $\chi^2$. However, for cooler stars, the $\chi^2$ statistic as a measurement of the model fit is less reliable due to features such as stellar flares that - as they are not included in the model atmosphere - can worsen the $\chi^2$ while the fundamental parameters are plausible. We discuss how we decide on the $\chi^2$ threshold in Section \ref{sec:mcmc_fits}.

\subsection{Priors}
\label{sec:mcmc_priors}
MaStar targets are first plotted in a CMD using the photometry values from the Gaia crossmatch. We use this to obtain prior estimates for T$_{\rm eff}$ and log g, with no estimate for [Fe/H]. In this instance, we overplot the PARSEC theoretical isochrone tracks \citep{bressan12} for ages from $2\;$Myr to $10\;$Gyr providing a fine grid and wide coverage in both parameters (we use $2,000-21,000\;$K in T$_{\rm eff}$ and $-6$ to $7.95$ in log g) and CMD space. Then, for each spectrum, we consider all isochrones that run through a box of $\Delta (\rm G_{BP}-G_{RP}) = 0.8$ and  $\Delta (\rm M_{G}) = 2$ taking the minimum and maximum values of T$_{\rm eff}$ and log g. For targets that fall outside of the isochrone coverage, we construct the same box around the closest isochrone point and follow the same procedure. Furthermore, we impose a minimum T$_{\rm eff}$ prior of at least $\pm16$ per cent of the nearest isochrone value for all targets as this allows for adequate sampling of $\pm500\;$K when analysing the coolest temperature range. The wide parameter coverage allows us to obtain priors for all parameter combinations and stellar types. 

\begin{figure*}
 \includegraphics[width=0.8\paperwidth]{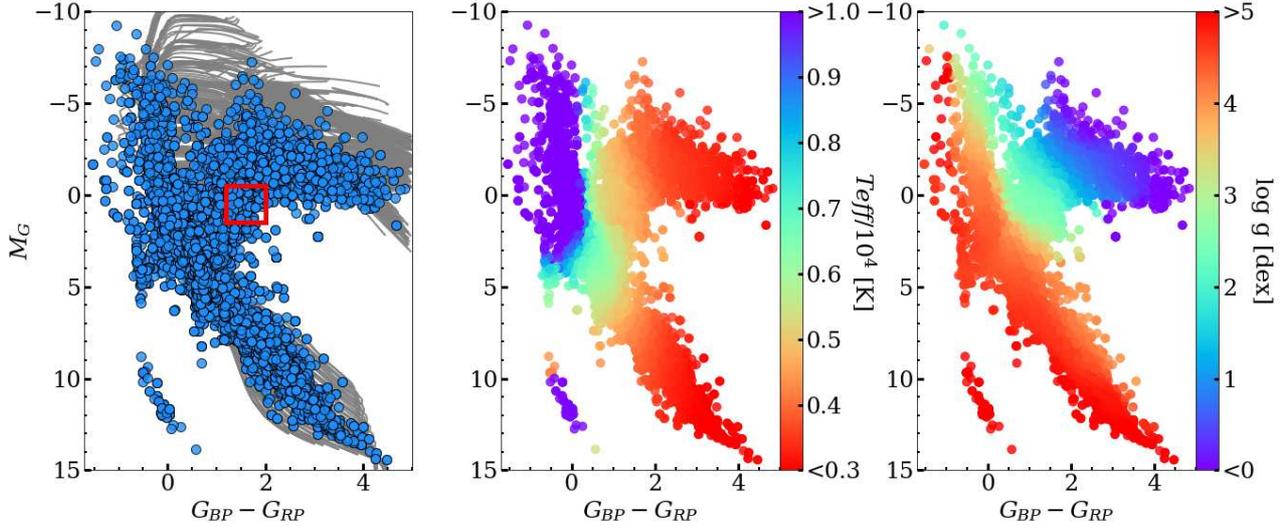}
 \caption{Left-hand panel: The coverage of PARSEC theoretical isochrones ($2\;$Myr to $10\;$Gyr) in grey over-plotted with the MaStar targets matched to Gaia photometry in blue. The PARSEC tracks allow for T$_{\rm eff}$ and log g estimates even for white dwarfs. The middle and right-hand panels show the resulting T$_{\rm eff}$ and log g priors, respectively.}
 \label{fig:isochrone_cmd}
\end{figure*}

The left-hand panel of Figure \ref{fig:isochrone_cmd} shows a CMD in Gaia colours with the PARSEC isochrones in grey, photometry-matched MaStar targets in blue and an example of the prior box in red. The middle and right-hand panels show the derived priors for T$_{\rm eff}$ and log g. We emphasise that these values represent the nearest isochrone values and that we take the minimum and maximum parameters within the predefined box which define a uniform prior over a fixed range.

\begin{figure*}
 \includegraphics[width=0.85\linewidth]{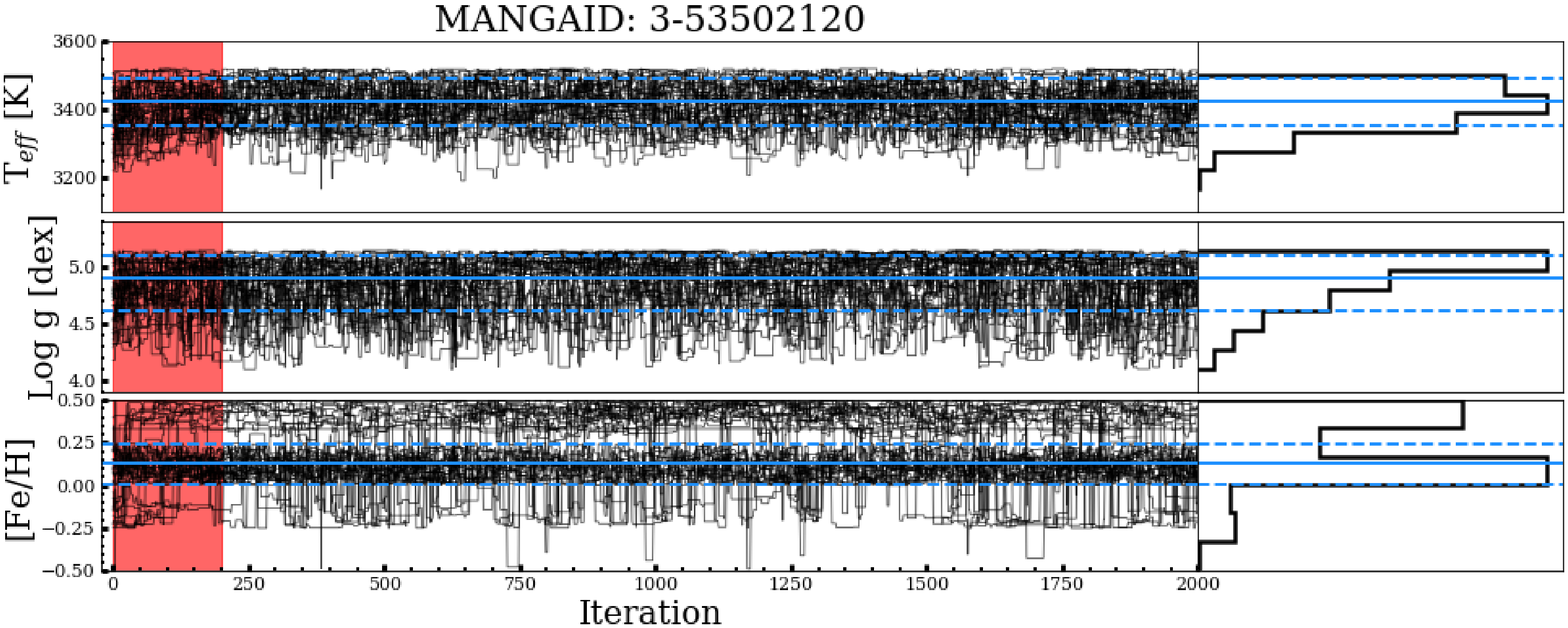}\hfill
 \includegraphics[width=0.85\linewidth]{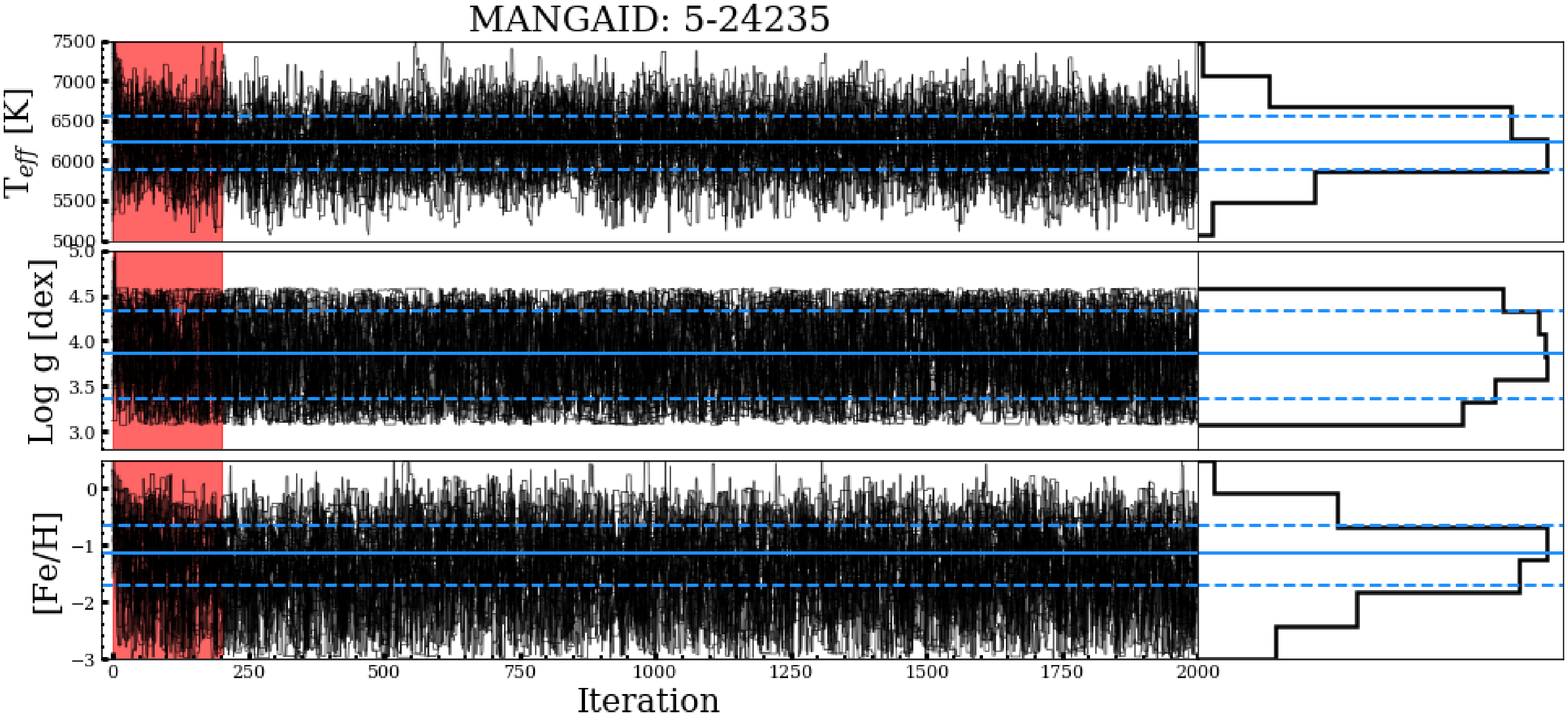}\hfill
 \includegraphics[width=0.85\linewidth]{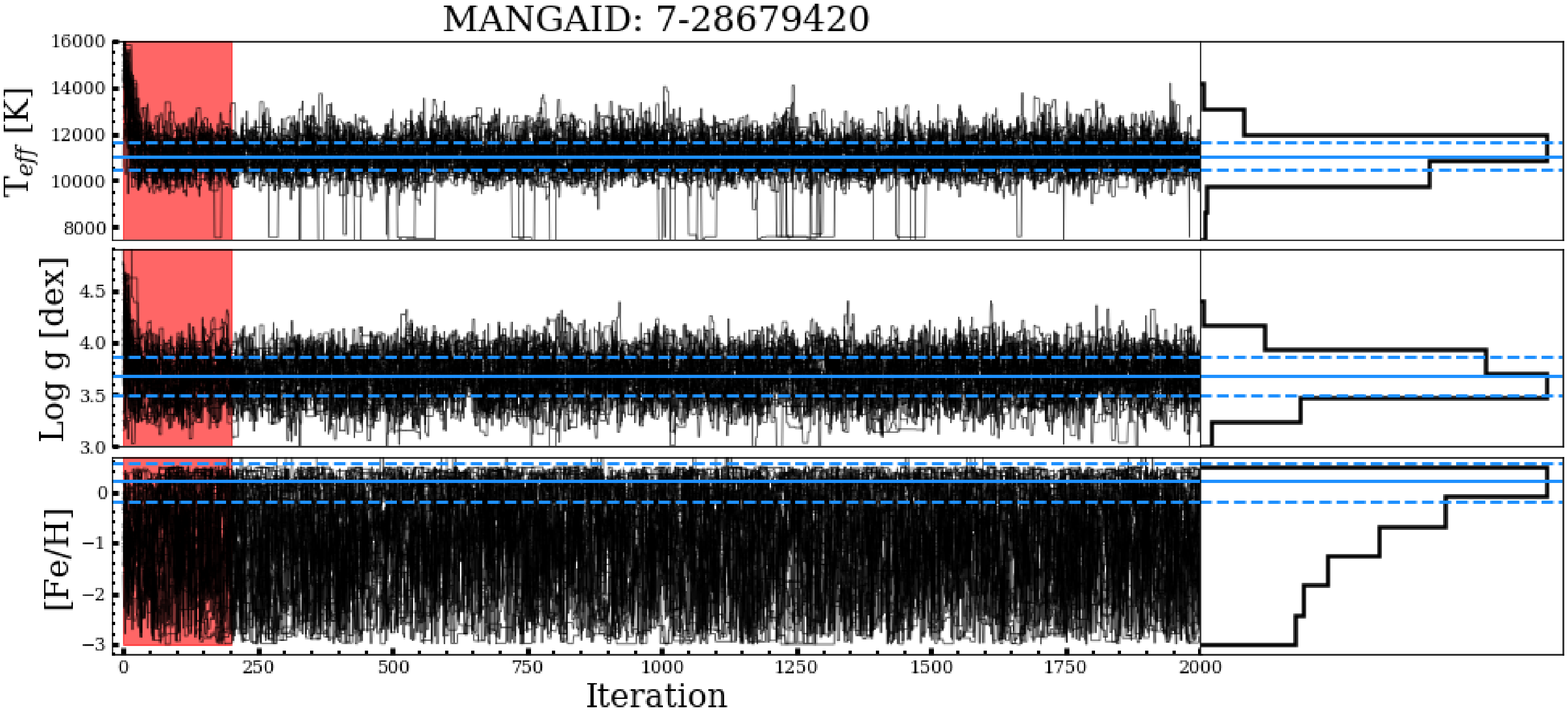}
 \caption{Example ensembles of walkers for the MCMC procedure. The top three panels show the how the FSPs are determined for a cool dwarf, the middle three for an intermediate temperature main sequence star and the bottom three for a hot main sequence star. The red shaded area of each panel shows the burn-in phase and the black lines represent the path of the walkers. The histogram for each panel shows the posterior distribution of the walkers, excluding the burn-in phase. The solid and dashed blue lines represent the selected parameter and one sigma errors, respectively. The priors for each target are as follows. MANGAID 3-53502120: $2824 \leq $T$_{\rm eff} \leq 3034$, $4.1 \leq $log g$ \leq 5.1$, $-3 \leq [{\rm Fe}/{\rm H}] \leq 0.5$. MANGAID 5-24235: $5069 \leq $T$_{\rm eff} \leq 8162$, $3.1 \leq $log g$ \leq 4.6$, $-3 \leq [{\rm Fe}/{\rm H}] \leq 0.5$. MANGAID 7-28679420: $6890 \leq $T$_{\rm eff} \leq 23356$, $2.6 \leq $log g$ \leq 4.7$, $-3 \leq [{\rm Fe}/{\rm H}] \leq 0.5$.}
 \label{fig:trace_3}
\end{figure*}

\begin{figure}
 \includegraphics[width=0.95\columnwidth]{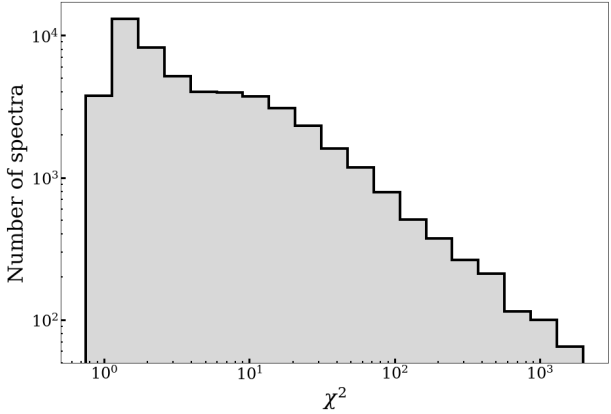}
 \caption{Distribution of $\chi^2$ values from the model fits.}
 \label{fig:chi_dist}
\end{figure}

\begin{figure}
 \includegraphics[width=0.95\columnwidth]{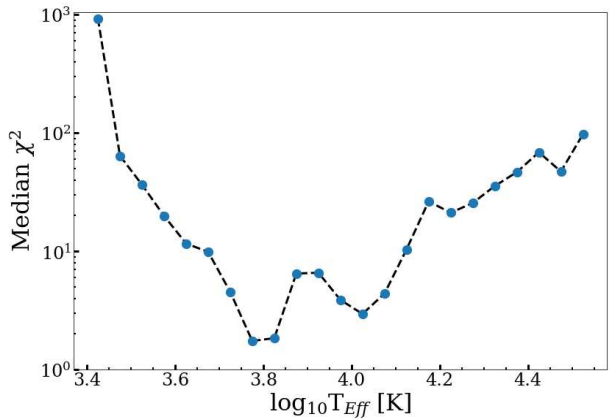}
 \caption{Median $\chi^{2}$ as a function of the binned effective temperature.}
 \label{fig:teff_vs_chi}
\end{figure}

\begin{figure*}
 \includegraphics[width=0.8\paperwidth]{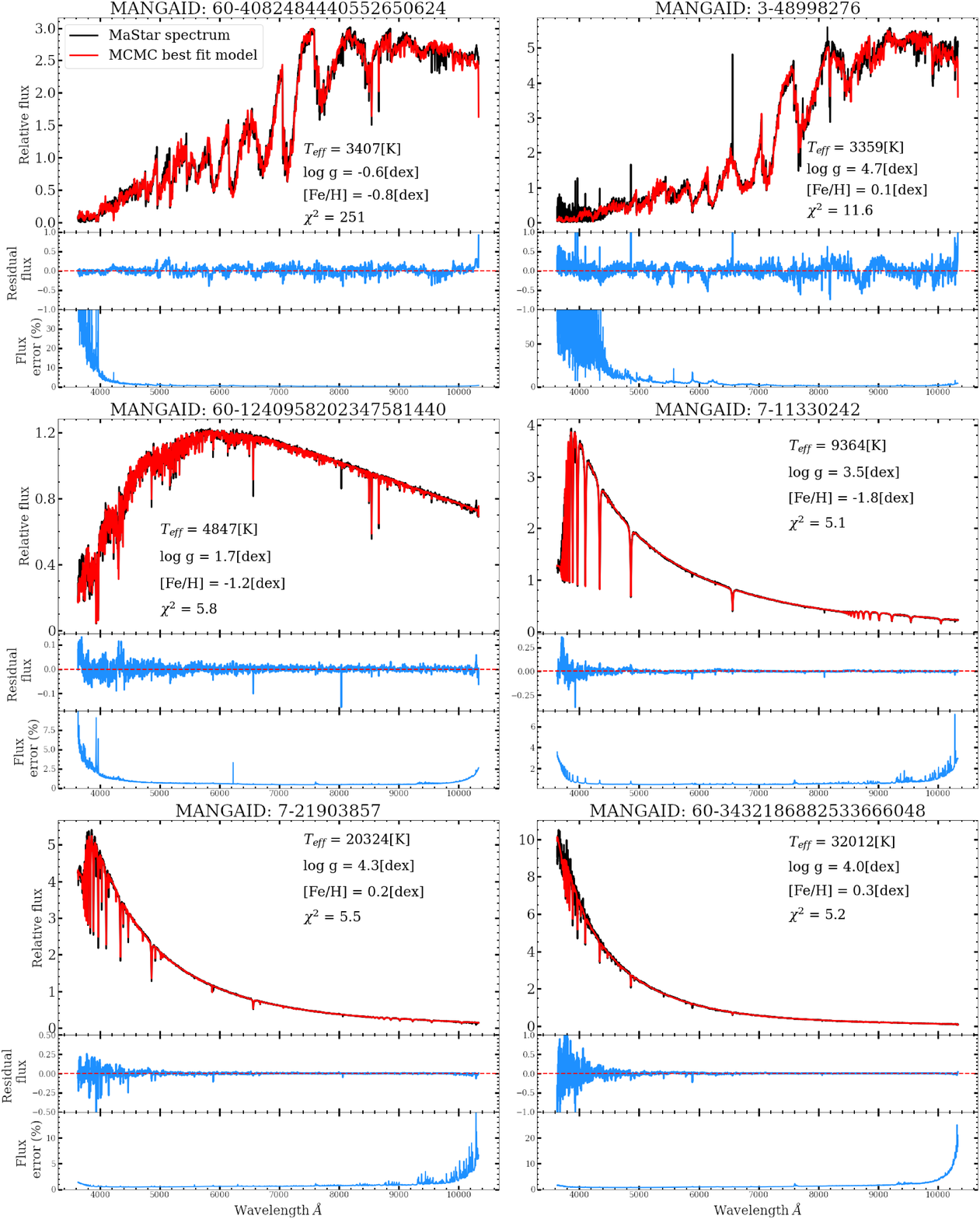}
 \caption{Example stellar fits and associated parameters using the MCMC full-spectral fitting method. Top panels: the extinction-corrected MaStar spectra and interpolated model fit. Middle panels: Absolute residual flux between the data and model. Bottom panels: The percentage error of the flux}
 \label{fig:mcmc_fits}
\end{figure*}

\subsection{Procedure}
The posterior distribution of each FSP is approximated using Baye's theorem:

\begin{equation}
    P(x|spec) = \frac{P(spec|x)P(x)}{P(spec)} \tag{2}
    \label{eq:2}
\end{equation}

where $x$ represents the model FSPs (T$_{\rm eff}$, log g and $[Fe/H]$) and \textit{spec} represents an individual, extinction-corrected spectrum. The posterior is defined by the probability ($P$) of the model given the observation and is represented by $P(x|spec)$. The likelihood is the probability of the observation given the model and is represented by $P(spec|x)$. The prior probability is defined by $P(x)$ in the absence of new data. $P({\rm spec})$ is ignored as this is the same division factor in all cases for the same observation. In other words, Equation \ref{eq:2} says that given the observed spectrum, the posterior probability of the parameters that are associated with the model $x$ can be found.

The log likelihood function is modelled as shown in equation \ref{eq:3} assuming that measurements are independent with Gaussian-distributed errors.

\begin{equation}
    P(spec|x) \propto -\frac{1}{2}\sum_{i=0}^{n} \left(\frac{spec_{i}-model_{i}(x)}{\sigma_{i}}\right)^2  \tag{3}
    \label{eq:3}
\end{equation}

The terms $spec_{i}$ and $\sigma_{i}$ represent the flux value and error at a given pixel in the spectrum. The term $model_{i}(x)$ represents the flux at each corresponding pixel of the linearly interpolated model spectrum. This is parameterised by the FSPs (\textit{x}) and has already been convolved by pPXF on the fly as each new model is proposed and used in the likelihood. The convolution involves fitting the proposed model with multiplicative polynomials of degree six. This allows one to account for small velocity offsets in the observed spectrum and to fit the spectral continuum in order to make a comparison between the observation and model. Sixth-degree multiplicative polynomials are used as this was found to produce the best balance between stable parameter recovery and fast computation speed.

\subsubsection{Affine Invariant Ensemble Sampler}
\label{sec:sampler}
Using MCMC, one is able to explore a multidimensional parameter space through the use of Markov chains - with each new proposal only being dependent on the last. The posterior of the target parameters can be sampled according to a number of available algorithms. In our analysis we use the Python package \texttt{emcee} \citep{emcee} which uses an implementation of the affine-invariant ensemble sampler proposed by \citet{goodman10}. We provide a brief overview of this algorithm, but refer the interested reader to the original paper. 

The property of affine-invariance allows the sampler to outperform standard samplers, such as the Metropolis-Hastings or Gibbs, when sampling anisotropic density distributions. This type of distribution is sometimes found in the relation between stellar FSPs. The sampler uses an ensemble of 'walkers' - stochastic chains of sampled points in each target parameter space - to probe the posterior shape of each parameter. The proposal distribution of each walker decides where it should move next in the chain. In this algorithm, the distribution is decided from the current positions of all other walkers in the ensemble (the complementary ensemble). A new position for a walker $X_{k}$ is proposed by randomly drawing a walker $X_{j}$ from the complementary ensemble. 

\begin{equation}
    X_{k}(t) \xrightarrow{} Y = X_{j} + Z[X_{k}(t) - X{j}]   \tag{4}
    \label{eq:4}
\end{equation}

$Y$ is the proposed position, $t$ represents a walker step and $Z$ is a random variable drawn from the distribution $g(z)$ which takes the following form \citep{goodman10}:

\begin{equation}
    g(z) \propto \begin{cases} 
    \frac{1}{z^{1/2}} & \text{if z} \in [\frac{1}{a}],a]\\
    0 & \text{otherwise}
    \end{cases}\tag{5}
    \label{eq:5}
\end{equation}

'\textit{a}' represents one of the few tuning parameters in this algorithm, which directly affects the acceptance fraction of walkers, \textit{i.e.} the number of proposed steps that are accepted. A suitable acceptance fraction allows the walkers to explore the parameter space without being trapped in local minima. \citet{emcee} suggest $a = 2$ and an acceptance fraction between $0.2 - 0.5$. We find this value of $a$ returns an acceptance fraction $>0.5$ and that $a = 5$ is more suitable for our pipeline, resulting in a mean acceptance fraction of 0.27 across all walkers. Other tunable parameters include the number of walkers and the number of steps in each walker chain.

\subsubsection{Parameter estimation}
To generate the posterior distribution we use $40$ walkers, each of $2,000$ steps, of which $200$ steps are discarded as a burn-in phase. The purpose of the burn-in is to remove any bias in the posterior from the starting positions of the walkers. Example trace plots of the ensemble walkers for three different MaStar targets of varying temperatures can be found in Figure \ref{fig:trace_3}. The burn-in period of 200 steps is shown by the red shaded area and the steps that form the posterior in the white area. The posterior distribution for each ensemble is shown by the histogram on the right of each panel, with the estimated parameters and errors shown by the solid and dashed horizontal lines respectively. The priors we use for each MANGAID are as follows. MANGAID 3-53502120: $2549 \leq $ T$_{\rm eff} \leq 3520\;$K, $4.1 \leq $ log g $ \leq 5.1\;$dex, $-3 \leq $ [Fe/{\rm H}] $ \leq 0.5\;$dex. MANGAID 5-24235: $5069 \leq $ T$_{\rm eff} \leq 8162\;$K, $3.1 \leq $ log g $ \leq 4.6\;$dex, $-3 \leq $ [Fe/{\rm H}] $\leq 0.5\;$dex. MANGAID 7-28679420: $6890 \leq $ T$_{\rm eff} \leq 23356\;$dex, $2.6 \leq $ log g $ \leq 4.7\;$dex, $-3 \leq$ [Fe/{\rm H}] $\leq 0.5\;$dex. These priors come from the CMD method described in Section \ref{sec:mcmc_priors}, where the prior range depends on the photometry of the MaStar target.

To avoid spurious effects caused by merging the two grids we explore both grids independently for cooler stars and use only BOSZ-ATLAS9 for hot stars. We found that BOSZ-ATLAS9 generally provide fits with better $\chi^2$ at these hotter temperatures. We use the minimum temperature of the T$_{\rm eff}$ prior to decide whether a star is to be analysed with both grids. If the minimum of the prior is greater than $5000\;$K, only the BOSZ-ATLAS9 models are used, otherwise both models are employed. In the case of both grids being used, the parameters are selected from the best fitting model with the best reduced $\chi^2$. 

We apply a small offset to the parameters in the RGB (log g $< 2$) that have been assigned parameters from MARCS models. The majority of spectra in this parameter space are covered by the grids of both models, except the few that have T$_{\rm eff} < 3500\;$K and the MARCS models come in to effect. We notice a small systematic offset (BOSZ-ATLAS9 subtract MARCS) of $-8\;$K in T$_{\rm eff}$, $0.08\;$dex in log g and $0.23\;$dex in [Fe/H] in the limits of $3500 \leq$ T$_{\rm eff} \leq 4100\;$K and log g $< 2$. We therefore add these offsets to MARCS based parameters in order to remove any bias. For the dwarf stars, we apply no offset as most of these can only be fit by MARCS models.

As the posterior distributions for T$_{\rm eff}$ and log g are produced by considering priors in the MCMC analysis, we take the median value of the posterior for these parameters. This is done to take into account any uncertainty in the distribution and to reflect the prior solution when no strong solution is found. For these parameters we calculate the errors using the $16^{th}$ and $84^{th}$ percentiles of the posterior. We estimate the best parameter for [Fe/H] by taking the maximum a posteriori (MAP), equivalent to the mode, as no prior is applied for this parameter and the MAP returns the best fitting parameter solution. To do this, a Gaussian kernel density estimation is used to estimate the probability density function and the maximum value of this returns the MAP. To estimate the errors for this parameter we take the $68$ percent credible interval and the corresponding [Fe/H] values at either end of the interval. 

Using the FSPs for each spectrum, we then produce an interpolated model spectrum and provide this to pPXF to obtain the polynomial corrected model that would have been used to evaluate the likelihood in the analysis. The reduced $\chi^2$ statistic for the observation and model is then calculated and used later for cleaning the final catalogue.

\subsubsection{Stellar fits}
\label{sec:mcmc_fits}

The distribution of $\chi^2$ values for each spectrum and model fit can be seen in Figure \ref{fig:chi_dist}. We also plot the relationship between log-binned T$_{\rm eff}$ and the median $\chi^2$ in Figure \ref{fig:teff_vs_chi}. As shown, there is a steep increase in $\chi^2$ for temperatures below $4000\;$K ($\log_{10}$T$_{\rm eff} = 3.6\;$K). We also note the increase in $\chi^2$ as temperatures increase above $12,500\;$K ($\log_{10}$T$_{\rm eff} = 4.1\:$K). As stars become bluer, the features in the spectrum become sparse, making it difficult to accurately fit models. Using these plots we decide to exclude FSP values with a $\chi^2 > 30$. Furthermore, we make an exception for spectra with T$_{\rm eff} < 4000\;$K and do not exclude these based on a $\chi^2$ criteria. This exception for poor fits at low temperatures is motivated by the complex features found at such temperatures combined with the shortcomings of theoretical atmospheres \citep{coelho07, gustafson08}. However, our method of fitting a wide wavelength range is less dependent on only a few lines that may carry large errors in synthetic spectra. Furthermore, the advantage of using a wide wavelength range which we explore with MaStar for the first time, is that we can use a large array of absorptions to constrain our parameters. Using this $\chi^2$ cut we are able to maintain 93 per cent of the data with Gaia values while removing poor fits. Combining this cut with the exclusion of data without distance estimates based on Gaia leaves 89 per cent of the $59,266$ spectra.

In Figure \ref{fig:mcmc_fits} we present model fits for a range of stellar types. For each example we show the model fit in the top panel, the residual fit in the middle and the flux error as a percentage of the flux in the lower panel. Even though modelling M type stars is generally challenging, we are able to fit both the TiO bands and the Ca II triplet features and recover relatively low $\chi^2$ values for a large proportion of these, as shown in the upper panels. The M giant spectrum in the upper left-hand panel is a fine example of why we allow high $\chi^2$ values for T$_{\rm eff} < 4000\;$K. Despite a good fit to the molecular bands and absorption lines, we recover a relatively high $\chi^2$ which is due to a large flux and small errors for most of this spectrum. We can also recover an excellent fit to the M dwarf spectrum shown in the upper right-hand panel despite a strong M-flare feature at $\sim6600\;$\AA{}. Furthermore, small residuals are also found for stellar types of higher temperature. Reliable fits of the models to the observations at different temperatures demonstrates how we can produce a robust set of parameters for calculating stellar population models. These excellent fits are supported by a median $\chi^2$ value of 2.3 for all spectra.

\begin{figure*}
 \includegraphics[width=0.9\paperwidth]{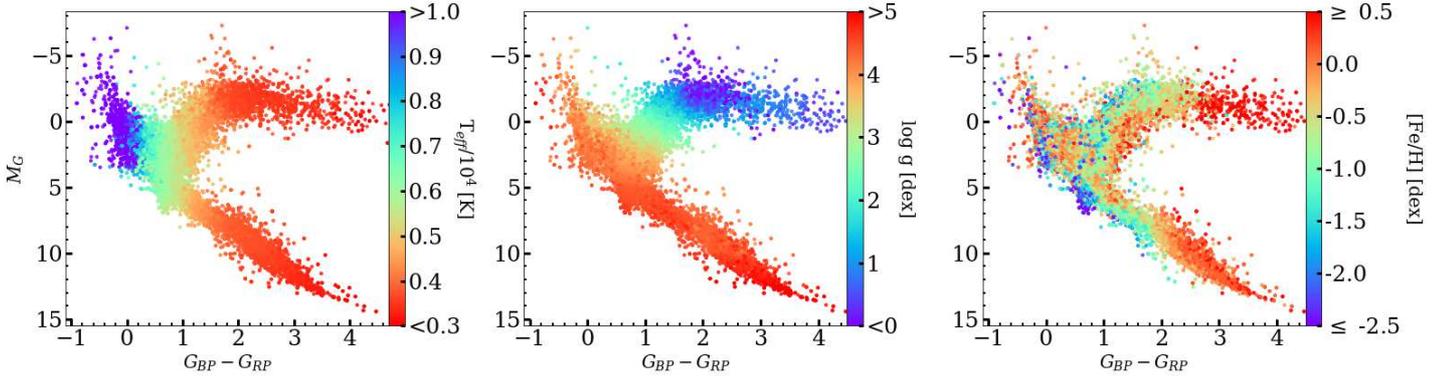}
 \caption{Extinction-corrected MaStar targets shown in a CMD with Gaia filters (G$_{BP}-$G$_{RP}$ vs. M$_{G}$), colour-coded by the recovered FSPs. The left-hand panel is colour-coded by T$_{\rm eff}$, the middle panel by log g and the right-hand panel by [Fe/H].}
 \label{fig:CMD_results_mpl9}
\end{figure*}

\begin{figure*}
 \includegraphics[width=0.65\paperwidth]{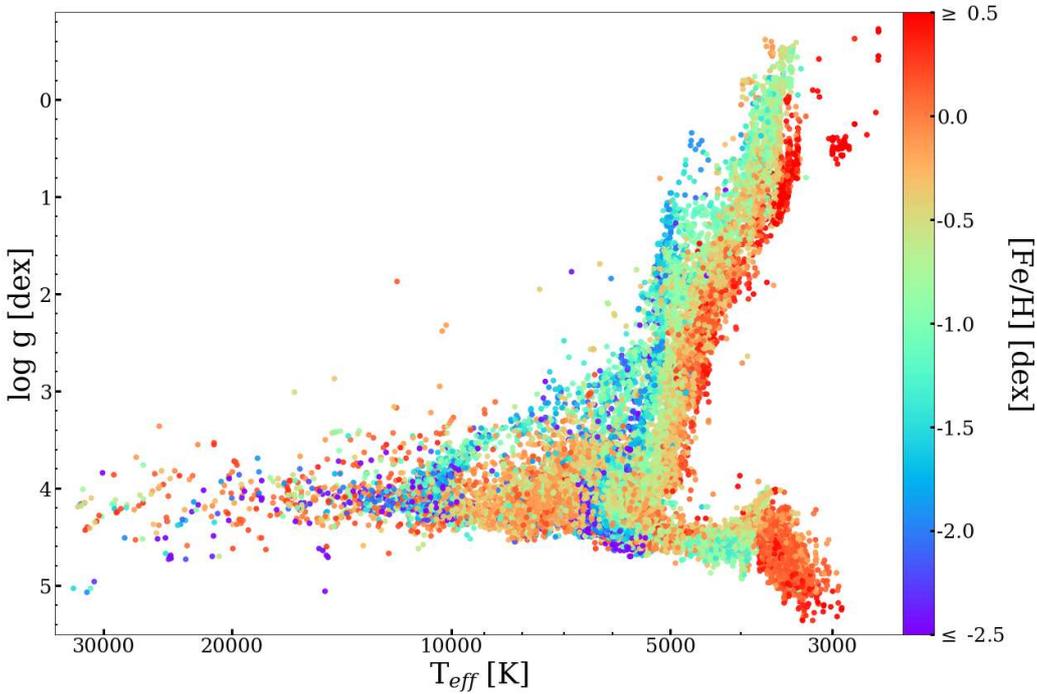}
 \caption{Spectroscopic Hertzsprung-Russel diagram showing the cleaned parameter distribution of MPL11 spectra, color coded by [Fe/H]. }
 \label{fig:HRD-v4}
\end{figure*}

\subsection{Results}
\label{sec:results-mcmc}

Presented in Figure \ref{fig:CMD_results_mpl9} are the CMDs for MaStar targets color-coded by the recovered parameters. We remove data belonging to white dwarfs, blue stragglers and EHB stars as well as maintaining the cut of $\chi^2 > 30$, except for stars with T$_{\rm eff} < 4000\;$K. This comparison also shows the improvement in coverage of spectral types in the MPL11 data release with respect to DR15 (MPL7) (Figure \ref{fig:CMD_results_mpl7}). In MPL11, an extension of dwarfs and giants to cooler temperatures, a hotter main sequence and a denser coverage of existing regions when compared to MPL7 is evident. We find T$_{\rm eff}$ to increase as the targets move to the bluer side of the CMD as expected. The minimum and maximum temperatures are $2592^{+1}_{-1}\;$K and $32983^{+4608}_{-2776}\;$K. The small error for the coolest star reflects that the spectrum is probably colder than what the model grid allows. Surface gravity also produces consistent results with no obvious outliers, decreasing in value towards the giant region of the plot. Blue horizontal branch stars that cross the main sequence are still present in the middle panel but are more difficult to see due to the density of points. The minimum and maximum values of log g are $-0.7^{+0.1}_{-0.1}\;$dex and $5.4^{+0.1}_{-0.1}\;$dex. [Fe/H] is more difficult to interpret, however for the main sequence and giant branches there is a decrease in [Fe/H] as G$_{BP}-$G$_{RP}$ decreases, leading to higher T$_{\rm eff}$. This is due to metal poor stars showing an ultraviolet excess due to line blanketing effects, which increases their T$_{\rm eff}$. The minimum and maximum values of [Fe/H] are $-2.9^{+0.4}_{-0.1}$ and $1.0^{+0.0}_{-0.1}\;$dex\footnote{Multiple spectra reach the maximum metallicity of the model grid, we present the average of their errors in this summary.}. Similar to the errors for the lowest temperature, these errors suggest that these spectra are likely more metal rich than the model grid allows.

In Figure \ref{fig:HRD-v4} we present the parameter distribution for all spectra with the previously described $\chi^2$ cut and Gaia $E(B-V)$ values in a HRD. This representation is powerful in that it allows one to easily identify outliers and mislabeled spectra. We find a well populated main sequence from the cool dwarf region through to the hotter side. There is a well-defined RGB structure in the range $3000 \leq $ T$_{\rm eff} \leq 5000\;$K and log g $ \leq 3$. For RGB stars with log g $< 1$, we note a preference to cooler temperatures. The complex features presented in such SEDs makes it difficult to obtain accurate T$_{\rm eff}$ values. The recovered RGB has a gradient in [Fe/H] moving to higher vales for cooler stars due to line blanketing effects. The cluster of metal rich stars at T$_{\rm eff} \approx 3000\;$K, log g $\approx 0.4\;$dex and the lower gravity stars scattered above are some of the reddest stars we observe in MaStar, with average Gaia colours (G$_{BP}-$G$_{RP}$) of 4.2. Therefore we are confident their temperature should place them separate from the main RGB, possibly even cooler \footnote{We suspect these are Oxygen rich variable stars and plan to further inspect them in future work.}. Cool dwarf stars at T$_{\rm eff} \leq 5000\;$K are also easily identified with high log g values extending down to $5.4\;$dex. The overall metallicity of the dwarf stars is typically metal rich. This distribution is consistent with other studies of local M-dwarf stars such as that by \citet{woolf-west12}. Furthermore, we can also recover parameters for intermediate stellar phases such as the sub giant branch at T$_{\rm eff} \geq 6000K$ and log g between 3 and 4 dex. The overall structure of this plot is consistent with our current understanding of stellar evolution as well as other HRDs in the literature.

In summarising the errors for each FSP, we split the T$_{\rm eff}$ parameter in to three bins of cool, intermediate and hot temperatures. In Table \ref{table:err_comp_mcmc} we show the errors for each FSP in temperature bins of T$_{\rm eff} < 5000\;$K, $5000 \geq $ T$_{\rm eff} < 15,000\;$K and T$_{\rm eff} \geq 15,000\;$K which are estimated from the PDF generated by the MCMC. We present the median of the upper and lower bound errors of each FSP. As the T$_{\rm eff}$ bins increase, there is an increase in the error of T$_{\rm eff}$ which is again due to fewer features in the SED of hotter stars and that these features become less sensitive to temperature changes. The uncertainty in log g is independent of T$_{\rm eff}$. We also see an increase in uncertainty for [Fe/H] at hotter temperatures. This is due to hotter stars having less absorption lines from metals which makes it difficult to accurately determine metallicity.

Parameter accuracy can also be interpreted from considering repeat observations of the same target. To do this, we calculate the standard error of each parameter, per target, and report the mean value of these errors. In Table \ref{table:err_comp_mcmc} the standard error of repeat observations from the cleaned catalogue are presented. For cool and intermediate temperatures the error in T$_{\rm eff}$ is similar with an increase for the the hottest stars, as seen in the PDF errors. Log g errors do not vary by much relative to their overall value and for [Fe/H], the error of the hottest stars is approximately three times larger which can be expected due to the same reason as the PDF error. The error from the PDF is larger than the standard error due to wide posteriors caused by subtle differences in the model spectra when estimating parameters to such precision. However, as the standard error is small for different observations of the same star, we demonstrate that the wide posterior doesn't cause a significant variation in parameters between observations.

\begin{table}
\caption{Median errors for each FSP with the MPL11 catalogue. Errors are split into low, medium and high temperature bins and are estimated two ways. Error source 'PDF' represents the errors calculated form the posterior generated by the walkers of the MCMC and 'Repeat obs' are calculated by taking the standard error of repeat observations.} 
\centering 
\begin{tabular}{c c c c c} 
\hline\hline 
FSP & Error & T$_{\rm eff}<\;$ & $5000 \geq $T$_{\rm eff}\;$ & T$_{\rm eff}\geq\;$ \\ [0.5ex]
 & source & 5000 & $<15,000$ & 15,000\\
\hline 
T$_{\rm eff}$ (K) & PDF & 166 & 413 & 2709 \\ 
 & Repeat obs & 12 & 22 & 296 \\
Log g (dex) & PDF & 0.4 & 0.4 & 0.4 \\
 & Repeat obs & 0.03 & 0.01 & 0.03 \\
$\rm[Fe/H]$ (dex) & PDF & 0.4 & 0.6 & 1.0 \\
 & Repeat obs & 0.04 & 0.07 & 0.22\\[1ex] 
\hline 
\end{tabular}
\label{table:err_comp_mcmc} 
\end{table}

\begin{figure*}
 \includegraphics[width=0.7\paperwidth]{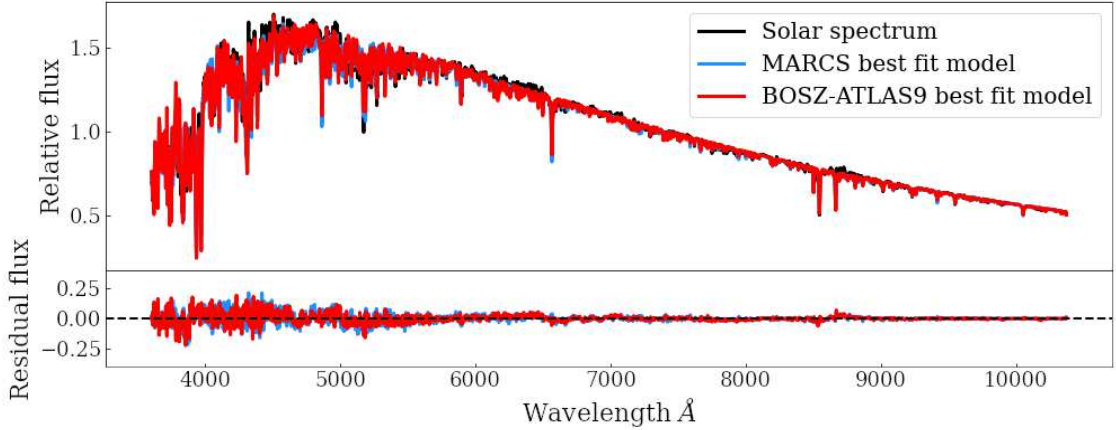}
 \caption{Full spectrum fit of the solar spectrum from \citet{meftah18} using our MCMC full-spectral fitting routine. \textit{Upper panel:} the black line represents the solar spectrum, the blue line represents the MARCS model with the selected parameters and the red line represents the BOSZ-ATLAS9 model with the selected parameters. \textit{Lower panel: } the residual of the solar spectrum subtract each model, with the same colour scheme as above. The black dashed line represents an exact fit to the data.}
 \label{fig:solar_fit}
\end{figure*}

\begin{figure*}
 \includegraphics[width=0.7\paperwidth]{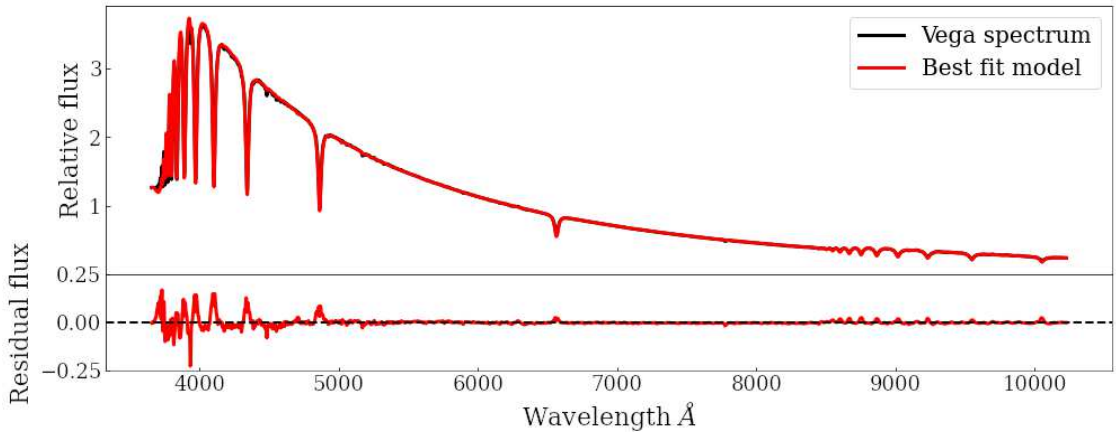}
 \caption{Full spectrum fit of Vega's spectrum from the CALSPEC archive \citep{bohlin14} using our MCMC full-spectral fitting routine. \textit{Upper panel:} the black line represents the data and the red line represents the BOSZ-ATLAS9 model with the selected parameters. \textit{Lower panel: } the residual of the data subtract the fitted model. The black dashed line represents an exact fit to the data.}
 \label{fig:vega_fit}
\end{figure*}

\begin{figure*}
 \includegraphics[width=0.7\paperwidth]{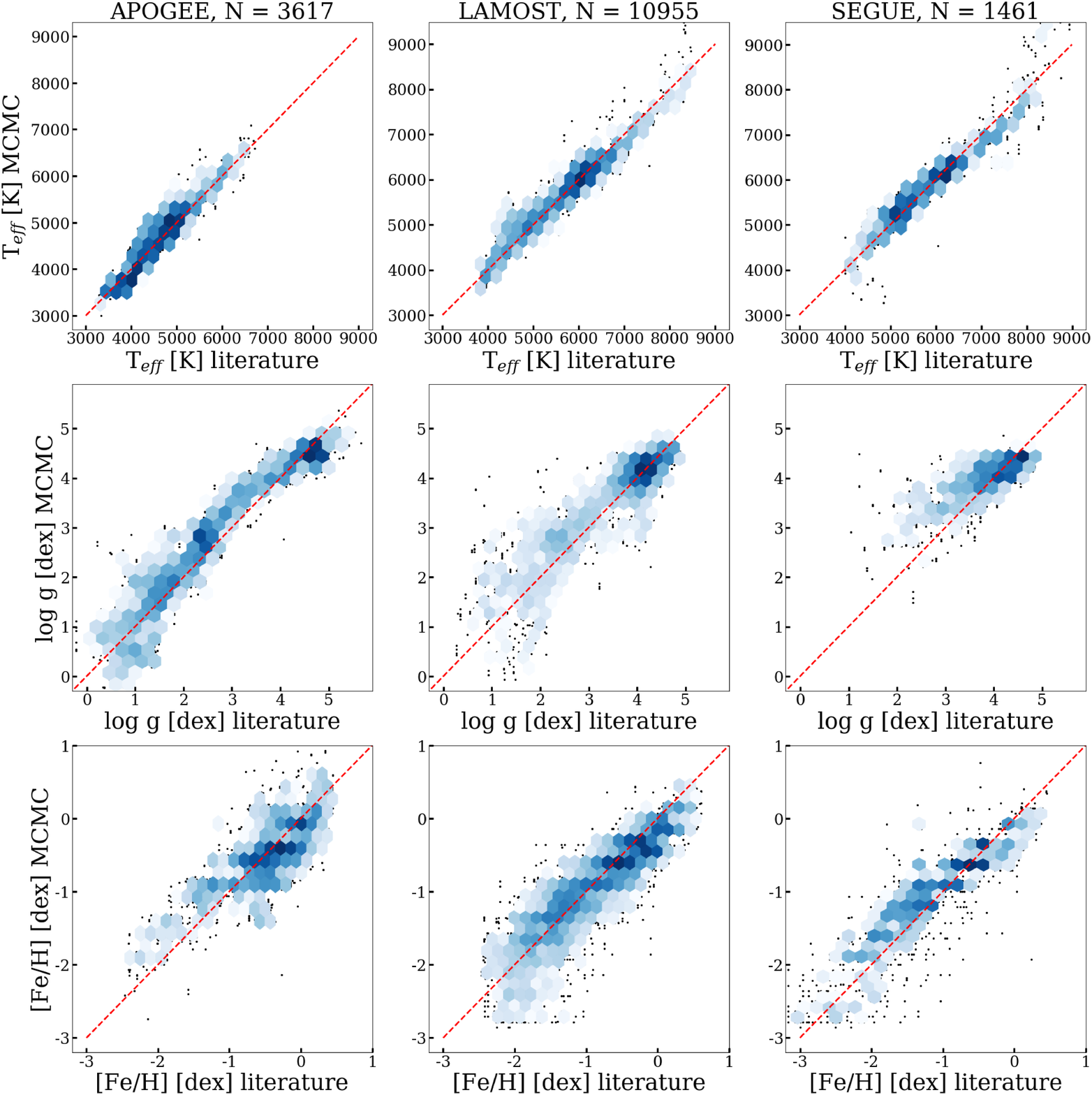}
 \caption{Comparison of our estimated parameters with APOGEE (left column), LAMOST (middle column) and SEGUE (right column) published parameters. Each panel shows the respective survey's parameters vs the parameters presented in this paper. The density of points is shown by the hexagons, where white is a minimum of five data points and dark blue shows higher density. Single black points are low density that aren't included in the minimum bin count of five. The red dashed line is a one-to-one marker that indicates perfect correlation.}
 \label{fig:mcmc_vs_lit}
\end{figure*}

\section{Comparison and testing}
\label{sec:comparison_testing}

\subsection{Stellar fits of the Sun and Vega}
By fitting the stellar atmospheres of nearby, standard stars, we are able to obtain a sense of the accuracy of our algorithm and use the results to calibrate our predictions. The benefit of using these nearby stars is that their atmospheres can be studied in great detail, leading to precise FSPs in the literature. For this analysis, we use the spectrum of the Sun obtained by the SOLar SPECtrometer (SOLSPEC, \citep{meftah18}) and the spectrum of Vega obtained by the Hubble Space Telescope, available in the CALSPEC archive\footnote{\url{http://www.stsci.edu/hst/observatory/crds/calspec.html}} \citep{bohlin14}. 

\subsubsection{Sun}
To determine the FSPs of the Sun, we use the MARCS and BOSZ-ATLAS9 models independently as the Sun's effective temperature is within the overlap region of the model grids. This was done to highlight any biases in either of the models and to avoid errors from non-continuous synthetic spectra at the MARCS and BOSZ-ATLAS9 interface. Despite the solar spectrum and models extending beyond the MaStar wavelength range ($3620 - 10350$ \AA), we fit the spectrum in this range in order to be consistent with our other results. Furthermore, we adopt no priors for this analysis and start the MCMC algorithm with an initial guess of T$_{\rm eff} = 5000\;$K, log g = $3\;$dex and [Fe/H] = $0\;$dex when analysing the solar spectrum.

In the upper panel Figure \ref{fig:solar_fit} we show the observed spectrum of the Sun and the full-spectral fit models from MARCS and BOSZ-ATLAS9. Provided with the SOL-SPEC file of the Sun was the uncertainty in the flux which was used in the likelihood of the MCMC and when calculating the reduced $\chi^2$ of the model fit. Here we see excellent modelling of the strong absorption features such as the NaD doublet at 5893\AA, H$\alpha$ at $6563$\AA\ and Ca II triplet between 8498 and 8662\AA. The quality of the model fit is also comparable between the two template sets; MARCS returns a reduced $\chi^{2} = 7.1$ and BOSZ-ATLAS9 a $\chi^{2} = 6.6$. The residuals of each fit are shown in the lower panel on the figure (observation subtract model). As one would expect from similar full-spectral fits, the recovered parameters are also comparable. The MARCS models recover T$_{\rm eff} = 5812^{+166}_{-186}\;$K, log g = $ 4.7^{+0.5}_{-0.5}$dex, $[Fe/H] = -0.3^{+0.3}_{-0.2}\;$dex. BOSZ-ATLAS9 models recover T$_{\rm eff} = 5771^{+136}_{-172}$, log g $ = 4.8^{+0.5}_{-0.6}$dex, $[Fe/H] = -0.1^{+0.3}_{-0.3}\;$dex. In \citet{meftah18}, they calculate the Sun's effective temperature by equating the integral of the solar spectrum to that of a blackbody spectrum at $5772\;$K. With the MARCS models we are 40K hotter than this and $1\;$K cooler with the BOSZ-ATLAS9 models, indicating that we recover T$_{\rm eff}$ well for this spectral type. The surface gravity of the Sun in log g is $4.437\;$dex \citep{sun_teff} which shows that for both models we slightly overestimate this parameter, but it's still within the quoted errors. Lastly, an underestimate of the metallicity [Fe/H] is found with both model sets, with the solar metallicity ([Fe/H] = $0\;$dex) within the estimated errors.

\subsubsection{Vega}
In Figure \ref{fig:vega_fit} we show the full-spectral fit of Vega's spectrum using the BOSZ-ATLAS9 models (MARCS models are not hot enough to fit this spectrum). We also take into account the uncertainty in the flux when calculating the FSPs here. As before, the top panel shows the data and model from the selected parameters, and the bottom panel shows the residual of the fit (observation subtract model). This fit returns a reduced $\chi^{2}$ of 5.8 which shows the flexibility of the interpolation in the MCMC method. The parameters recovered are: T$_{\rm eff} = 9134^{+780}_{-728}\;$K, log g = $ 4.0^{+0.3}_{-0.3}\;$dex, $[Fe/H] = -1.2^{+1.2}_{-1.2}\;$dex. In \citet{castelli94}, they cite that for Vega: T$_{\rm eff} = 9550\;$K, log g $= 3.95\;$dex and $[Fe/H] = -0.5\;$dex. For T$_{\rm eff}$, we find a value with relatively small uncertainty and consistent with the literature within our margin of error. We are also able to recover a very accurate value for log g with small uncertainty. Finally, our value of [Fe/H] is lower than the literature by 0.7 dex. There is however a considerable uncertainty in our estimate of this parameter. The metallicity of hot stars is difficult to determine as they have fewer absorption features that can be fit. As a result, we find a relatively flat posterior at low metallicities for Vega.

\subsection{Comparison with literature}
\label{sec:lit-comp}

\begin{table*}
\caption{Statistical comparison of our parameters from the MCMC procedure with literature values of APOGEE, SEGUE and LAMOST. We present the median of the differences between the two sets (this work subtract literature) and the standard deviation ($\sigma$) of the difference for three stellar phases: Dwarfs (T$_{\rm eff} < 5000$ \& log g $\geq4.5$), RGB (T$_{\rm eff} < 6000$ \& log g $< 4.0$) and MS/HB (all remaining spectra).} 
\centering 
\begin{tabular}{c c c c c c c c c c c} 
\hline\hline 
Source & Parameter & Median difference & $\sigma$ & Median difference & $\sigma$ & Median difference & $\sigma$ \\ [0.5ex]
 & & (Dwarf) & (Dwarf) & (RGB) & (RGB) & (MS/HB) & (MS/HB)\\ [0.5ex]
\hline 
APOGEE & T$_{\rm eff}$ (K) & -161 & 127 & 39 & 202 & -37 & 227 \\ 
 & Log g (dex) & -0.1 & 0.2 & 0.3 & 0.4 & -0.1 & 0.3 \\
 & [Fe/H] (dex)& -0.3 & 0.5 & 0.0 & 0.3 & -0.1 & 0.3 \\
LAMOST & T$_{\rm eff}$ (K) & -18 & 118 & 54 & 217 & -47 & 146 \\
 & Log g (dex) & -0.1 & 0.2 & 0.2 & 0.6 & 0.0 & 0.2 \\
 & [Fe/H] (dex) & -0.3 & 0.3 & 0.0 & 0.3 & -0.1 & 0.3 \\
SEGUE & T$_{\rm eff}$ (K) & -94 & 377 & 142 & 204 & -49 & 295 \\ 
 & Log g (dex) & 0.0 & 0.2 & 0.3 & 0.6 & 0.0 & 0.4 \\
 & [Fe/H] (dex) & -0.1 & 0.5 & 0.3 & 0.3 & 0.1 & 0.4 \\[1ex] 
\hline 
\end{tabular}
\label{table:lit_comp} 
\end{table*}

Due to the target selection strategy used for MaStar, there is a large overlap between stars with literature values and MaStar observations. Using the Two Micron All Sky Survey (2MASS, \citet{2mass06}) IDs of MPL11 stars we find 2088 unique matches (4786 matches including repeat observations) with APOGEE/APOGEE-2N DR16 parameters \citep{apogee-dr16} that have been derived using ASPCAP. To compare with the LAMOST DR6-v2 A,F,G,K star catalogue parameters\footnote{\url{http://dr6.lamost.org/v2/}}, derived using the LAMOST Stellar Parameter pipeline (LASP, \citet{LASP2015}), we match MPL11 stars based on the RA and Dec, with a maximum error of 0.0001 deg. Doing so, we find 4954 unique matches and 12,721 matches to repeat observations. For the comparison with SEGUE we use the cross match from the initial targeting for observations, as described in Section \ref{sec:observations}. These parameters have been derived using the SEGUE Stellar Parameter Pipeline (SSPP, \citet{Lee:2007mf, Lee:2007ec}). This returns 804 unique matches to MPL11 stars and 2457 matches to all observations. 

For the three literature catalogues described we clean the matches based on our reduced $\chi^2$ criteria of 30, as previously described. In Figure~\ref{fig:mcmc_vs_lit} we show a one-to-one comparison of each survey and the corresponding parameters we recover. At the top of each column are the number of observations that are compared. The hexagons show the density of points, the darker shade representing denser areas, with black markers representing single spectra. The red dashed line shows the fiducial one-to-one line for comparison. 

Considering the T$_{\rm eff}$ comparison (top row), we see a small systematic offset for SEGUE and LAMOST at temperatures above $\sim6500$ K, where we are underestimating the temperature compared to the literature. Slight systematic trends are seen for log g. APOGEE has the most homogeneous distribution of log g values, from which we tend to overestimate at values below $4\;$dex. We also overestimate compared to LAMOST at intermediate values, with increasing scatter towards lower gravity. For all catalogues we are consistent between the densest region of $4 <$ log g $< 5\;$dex. For [Fe/H], the LAMOST comparison shows that we find more metal poor solutions where they find values of $-2 <$ [Fe/H] $< -1.5\;$dex. This may be due to the LASP having few examples of the most metal poor stars as they rely on empirical templates. Apart from this, there is good agreement with APOGEE and LAMOST, with some scatter. We also note a slight systematic offset with the SEGUE comparison between $-2$ and $-1$ dex. 

\begin{table}
\caption{Median errors for the spectra compared to APOGEE, LAMOST and SEGUE in Table \ref{table:lit_comp}. As done before, we split this analysis in to three stellar phases: Dwarfs, RGB and MS/HB stars.} 
\centering 
\begin{tabular}{c c c c c} 
\hline\hline 
Source & Parameter & Error & Error & Error \\ [0.5ex]
 &  & (Dwarf) & (RGB) & (MS/HB) \\ [0.5ex]
\hline 
APOGEE & T$_{\rm eff}$ (K) & 152 & 225 & 167\\ 
 & Log g (dex) & 0.3 & 0.6 & 0.3 \\
 & [Fe/H] (dex)& 0.4 & 0.4 & 0.4 \\
LAMOST & T$_{\rm eff}$ (K) & 175 & 256 & 353 \\
 & Log g (dex) & 0.3 & 0.6 & 0.4 \\
 & [Fe/H] (dex) & 0.4 & 0.4 & 0.6 \\
SEGUE & T$_{\rm eff}$ (K) & 281 & 430 & 475 \\ 
 & Log g (dex) & 0.3 & 0.6 & 0.4 \\
 & [Fe/H] (dex) & 0.6 & 0.7 & 0.8 \\[1ex] 
\hline 
\end{tabular}
\label{table:lit_comp_errors} 
\end{table}

The median and standard deviation of the differences between our values and the literature are presented in Table \ref{table:lit_comp}, where we also distinguish between three key stellar phases. We define these phases in T$_{\rm eff}$ and log g space based on our parameters: dwarfs (T$_{\rm eff} < 5000\;$K, log g $\geq 4.5$), RGB (T$_{\rm eff} < 6000\;$K, log g $< 4$), MS and HB (remaining spectra). In general we obtain a lower T$_{\rm eff}$ for dwarfs and MS/HB and larger for RGB stars. Overall we find the smallest offset to all three surveys when estimating T$_{\rm eff}$ for the MS/HB. The dispersion seems to be most significant for SEGUE across all stellar types, with the one towards LAMOST being the smallest on average. This is interesting as the latter parameters are also based on fitting a medium-resolution spectrum over a wide wavelength range.

The median difference and dispersion in log g is the largest for each survey when comparing against the RGB stars.
This could be due to the RGB containing the largest range in log g for the three stellar phases we define. For dwarfs and the MS/HB, no significant offset is found for each survey. The scatter for log g is larger for the LAMOST and SEGUE RGB comparison, but as shown by Figure \ref{fig:mcmc_vs_lit}, the statistics are based on only a few spectra due to a lack of observations by the literature in this parameter space.

For metallicity the systematic offsets are small, for the APOGEE and LAMOST RGB comparison we find no offset. Except for the RGB, we find slightly larger dispersions for SEGUE parameters.

In Table \ref{table:lit_comp_errors} we compare the median errors we calculate from the posterior of each parameter with the same sample stars in each literature comparison for the three stellar phases. Comparing these errors to the dispersion values in Table \ref{table:lit_comp} we can get a sense of whether our errors are sensible. For T$_{\rm eff}$, the errors we find are approximately equal to the APOGEE dispersion, with the largest discrepancy being in MS/HB stars where our error are lower by $60\;$K. This is similar for LAMOST, however our errors in the MS/HB are larger by $207\;$K. Errors differ the most for the SEGUE comparison, where ours are larger for the RGB and MS/HB by $226$ and $180\;$K, respectively. The error for log g is approximately equal to the dispersion of all surveys and all stellar phases. Importantly, the increase in dispersion for RGB stars we see across all surveys is reflected by a larger error in our parameters of $0.6\;$dex. Comparing with APOGEE, our errors in [Fe/H] differ from the dispersion by $0.1\;$dex. For LAMOST MS/HB stars our errors on [Fe/H] are larger by $0.3\;$dex. Our errors are also larger for SEGUE RGB and MS/HB stars.

\section{SSP models as output and as a test}
\label{sec:ssp}
As discussed in the Introduction, stellar parameters are needed for including empirical stellar spectra in a population synthesis model. Consequently, stellar population models also serve as a guide to identify areas requiring further calibration of the input stellar library. This approach was put forward in \citetalias{maraston20} for the models obtained with the discrete $\chi^2$ Method (Appendix \ref{sec:method1}) on the first MaStar data release and is being adopted also for the parameters generated for MPL11 with the MCMC Method. 

The stellar parameters calculated with the discrete $\chi^2$ Method as well as those by Chen et al. (2020) were used in \citetalias{maraston20} to generate two sets of stellar population models, named \textit{Th-MaStar} and \textit{E-MaStar}, respectively \footnote{available at \url{https://www.icg.port.ac.uk/mastar/}}, covering intermediate/old ages $0.1 - 15\;$Gyr, metallicities ([\textit{Z}/H]) from $-2.35$ to $0.35\;$dex, for various IMFs. As described, the discrete $\chi^2$ Method is referred to as \textit{Th} due to the adoption of purely theoretical stellar atmospheres as a reference, whereas the method of Chen et al. is referred to as \textit{'E'} as they use a semi-empirical reference library. Specifically, empirical spectra are used for T$_{\rm eff} < 20,000\;$K and theoretical templates otherwise. Furthermore, chemical composition measurements are ultimately derived using theoretical model atmospheres.

In \citetalias{maraston20}, a thorough comparison of the \textit{E} and \textit{Th} models and the conclusions we could draw on the respective stellar parameters is carried out and we refer the reader to the paper for more detail. Here we briefly summarise their findings, as they are relevant to the stellar parameter calculation we perform in this paper.

Firstly, the \textit{Th} models are able to probe to lower ages than the \textit{E} models due to the wider parameter coverage. Particularly in the very metal-poor regime ($[Z/{\rm H}] = -2.25$), the \textit{Th} parameters allow for model creation with main sequence turn-off ages down to 1 Gyr, compared to ages of 6 Gyr with the \textit{E} parameters (see Table 1 in \citetalias{maraston20}). With the \textit{Th} parameters we are also able to identify low-mass main sequence dwarfs down to the hydrogen burning limit. Similar results are found for the metal-poor regime ($[Z/{\rm H}] = -1.35$), with \textit{Th} and \textit{E} models reaching turn-off ages as low as 0.5 Gyr. The \textit{Th} models show a good match between the RGB slope and what theoretical isochrones suggest, while the RGB of the \textit{E} models is slightly cooler for solar and half-solar $[Z/{\rm H}]$. Furthermore, for the old, metal-rich populations, the \textit{Th} models can reproduce significant near-IR bands thanks to the identification of cool upper-RGB spectra.

To further test the accuracy of population models and stellar parameters, \citetalias{maraston20} derived ages and metallicity for observed star cluster spectra through full spectral fitting \citep{firefly17} and compared with literature values. The result was that metallicity estimates are similarly well reproduced for both models, with a median offset of approximately $-0.1\;$dex and a scatter of $0.6\;$dex, while ages were better reproduced by the \textit{Th}-models, leading to offsets that could be as small as 9 per cent, in dependence of the assumed literature values. 

The results from \citetalias{maraston20} support our use of only theoretical spectra and full spectral fitting for the calculation of stellar parameters, which led us to expand our procedure into the MCMC Method. Although Chen at al. parameters ultimately rely on theoretical atmospheres, the main advantage of our \textit{Th}-models is that we are not restricted to the grid of the reference empirical library. The stellar parameters calculated with the MCMC Method have been subjected to the same procedure, namely test population models were calculated in parallel to the parameter calculations in order to check the accuracy of the parameters, any gap in evolutionary phase coverage, the exact cut to apply in chi-square, etc. Stellar population models based on the parameters from the MCMC Method are now able to extend to much younger ages and will be published in a fortcoming paper (Maraston et al. {\it in prep.})

\section{Conclusions}
\label{sec:conclusion}
The MaNGA Stellar Library (MaStar) \citep{yan19} will be the largest library of stellar spectra upon completion. By using the MaNGA IFU fiber bundles and software pipeline with the BOSS spectrographs, MaStar has the same wavelength range and wavelength dependent resolution as MaNGA spectra. Furthermore, the catalogue contains a wide variety of stellar types which allows for the creation of robust stellar population models \citep{maraston20}. 

We expand our method for stellar parameter calculations \citep{maraston20} based on fitting theoretical spectra from model atmospheres with the full spectral fitting code pPXF \citep{cappellari04, ppxf17} by adding an MCMC procedure for fully mapping the parameter space and quantifying uncertainties. We also define a more elaborated method for including constraints from GAIA photometry. We apply this method to the complete ensemble of MaStar, consisting of 59,266 per-visit-spectra for 24,290 unique stars to determine the fundamental atmospheric stellar parameters (T$_{\rm eff}$, log g and [Fe/H]). Using Gaia matched photometry and theoretical isochrones we are able to first set realistic priors on the T$_{\rm eff}$ and log g parameters before computing their posteriors. FSPs are then determined by comparing observations to synthetic stellar atmospheres of MARCS \citep{gustafson08} and BOSZ-ATLAS9 \citep{meszaros12, bohlin17}. By using the full wavelength range of MaStar we are less sensitive to individual and potentially spurious absoption lines in the synthetic spectra. Using this approach, we are able to obtain accurate full-spectral fits across all spectral types which is crucial for creating stellar population models. We find a median $\chi^2$ of 2.3 for spectra with recovered parameters. 

The outcome is a comprehensive parameter catalogue, spanning $2592^{+1}_{-1} \leq $ T$_{\rm eff} \leq 32983^{+4608}_{-2776}\;$K; $-0.7^{+0.1}_{-0.1} \leq $ log g $ \leq 5.4^{+0.1}_{-0.1}\;$dex; $-2.9^{+0.4}_{-0.1} \leq [{\rm Fe}/{\rm H}] \leq 1.0^{+0.0}_{-0.1}\;$. Uncertainties are also provided. The corresponding H-R diagram has a sound overall structure, with well defined branches and a metallicity structure obeying stellar evolution.

For testing our procedure, we perform full-spectral fits of the Sun (Figure \ref{fig:solar_fit}) and Vega (Figure \ref{fig:vega_fit}) in order to test our recovered parameters against very well known, empirically derived, stellar parameters. Using the MARCS and BOSZ-ATLAS9 models independently, we find excellent results for both calibration stars.

Our procedure is further tested by comparing with the literature parameters of independent surveys obtained with their respective stellar parameter pipelines. This comparison is split in to three key stellar phases: dwarfs, RGB and MS/HB stars. We find the closest agreement in terms of median offset and dispersion with LAMOST and APOGEE across the three stellar types. This is encouraging as LAMOST offers the largest cross match of data and therefore represents a more statistically significant result and APOGEE is based on high-resolution spectroscopy on a small and completely independent wavelength range. For log g, the RGB showed the largest dispersion with LAMOST and SEGUE which may result from this stellar phase containing the largest range in log g values and the fewest number of cross matched stars. Overall the statistics show a good correlation for [Fe/H] except for SEGUE.

As there are many steps in each pipeline, some disagreement between catalogues is expected and it is difficult to conclude on which represents the 'true' atmospheric parameters. To probe the correctness of our parameters, we ultimately use stellar population models based on the parameters and assess how well they are able to recover independent age and metallicity values of stellar systems, as we have done in \citet{maraston20}. In that paper we showed that - even with the relatively simple discrete $\chi^2$ method (see Appendix \ref{sec:method1}) - we are able to obtain a set of stellar parameters that allows us to create accurate stellar population models which could reproduce well the ages and metallicities of intermediate-age and old star clusters. With the full MaStar catalogue and the extended MCMC method we are now able to calculate stellar population models down to a few Myr of age (Maraston et al. in prep.), which we shall probe against globular cluster fitting in a fortcoming paper.

As a final remark, in this work we have considered only the three fundamental atmospheric stellar parameters required for creating stellar population models. In Appendix \ref{sec:additional_params} we briefly explore the effect of changing rotational velocity and microturbulence in the model grids and found both effects to be negligible at MaStar's spectral resolution. We now plan to derive individual element abundances for the MaStar spectra.

\section*{Acknowledgements}
LH acknowledges support from the Science \& Technology Facilities Council (STFC) as well as the Data Intensive Science Centre in SEPnet (DISCnet). STFC is also acknowledged by JN for support through the Consolidated Grant Cosmology and Astrophysics at Portsmouth, ST/S000550/1. RY and DL acknowledge support by NSF grant AST-1715898.
Numerical computations were done on the Sciama High Performance Compute (HPC) cluster which is supported by the ICG, SEPNet and the University of Portsmouth.

Funding for the Sloan Digital Sky 
Survey IV has been provided by the 
Alfred P. Sloan Foundation, the U.S. 
Department of Energy Office of 
Science, and the Participating 
Institutions. SDSS-IV acknowledges support and resources from the Center for High 
Performance Computing  at the 
University of Utah. The SDSS 
website is www.sdss.org.

SDSS-IV is managed by the 
Astrophysical Research Consortium 
for the Participating Institutions 
of the SDSS Collaboration including 
the Brazilian Participation Group, 
the Carnegie Institution for Science, 
Carnegie Mellon University, Center for 
Astrophysics | Harvard \& 
Smithsonian, the Chilean Participation 
Group, the French Participation Group, 
Instituto de Astrof\'isica de 
Canarias, The Johns Hopkins 
University, Kavli Institute for the 
Physics and Mathematics of the 
Universe (IPMU) / University of 
Tokyo, the Korean Participation Group, 
Lawrence Berkeley National Laboratory, 
Leibniz Institut f\"ur Astrophysik 
Potsdam (AIP),  Max-Planck-Institut 
f\"ur Astronomie (MPIA Heidelberg), 
Max-Planck-Institut f\"ur 
Astrophysik (MPA Garching), 
Max-Planck-Institut f\"ur 
Extraterrestrische Physik (MPE), 
National Astronomical Observatories of 
China, New Mexico State University, 
New York University, University of 
Notre Dame, Observat\'ario 
Nacional / MCTI, The Ohio State 
University, Pennsylvania State 
University, Shanghai 
Astronomical Observatory, United 
Kingdom Participation Group, 
Universidad Nacional Aut\'onoma 
de M\'exico, University of Arizona, 
University of Colorado Boulder, 
University of Oxford, University of 
Portsmouth, University of Utah, 
University of Virginia, University 
of Washington, University of 
Wisconsin, Vanderbilt University, 
and Yale University.

\section*{Data Availability}
We shall release the data underlying this article only through the official releases of the SDSS-IV.

\addcontentsline{toc}{section}{Acknowledgements}



\bibliographystyle{mnras}
\bibliography{ms.bbl}

\appendix

\section{Discrete \texorpdfstring{$\chi^2$}{k} Method}
\label{sec:method1}
Here we recap the method for parameter determination employed to create MaStar intermediate - old age stellar population models in \citetalias{maraston20}. This method was applied to the DR15 (MPL7) MaStar spectra - 8,646 spectra for 3,321 unique stars. This method has now been expanded as described in the paper and used from MPL11 onward.

\subsection{Synthetic Stellar Atmospheres and Priors}
\label{sec:chi_priors}
We compare observations to theoretical spectra from the model atmospheres of MARCS \citep{gustafson08} and BOSZ-ATLAS9 \citep{meszaros12, bohlin17}. The combined model grid can be seen in Figure \ref{fig:marcs-atlas_grid}. In the area where the grids overlap, they are explored simultaneously and the FSPs are then drawn from the best fitting template. In this common region we find $\sim 88 \%$ of the observations matched to BOSZ models and the remaining matched to MARCS models.

Before analysis, we narrow the theoretical grid by adopting priors based on photometry to reduce computation time. We have verified that the result without the application of these priors is the same, but the calculation time is much longer. 

\begin{figure}
 \includegraphics[width=\columnwidth]{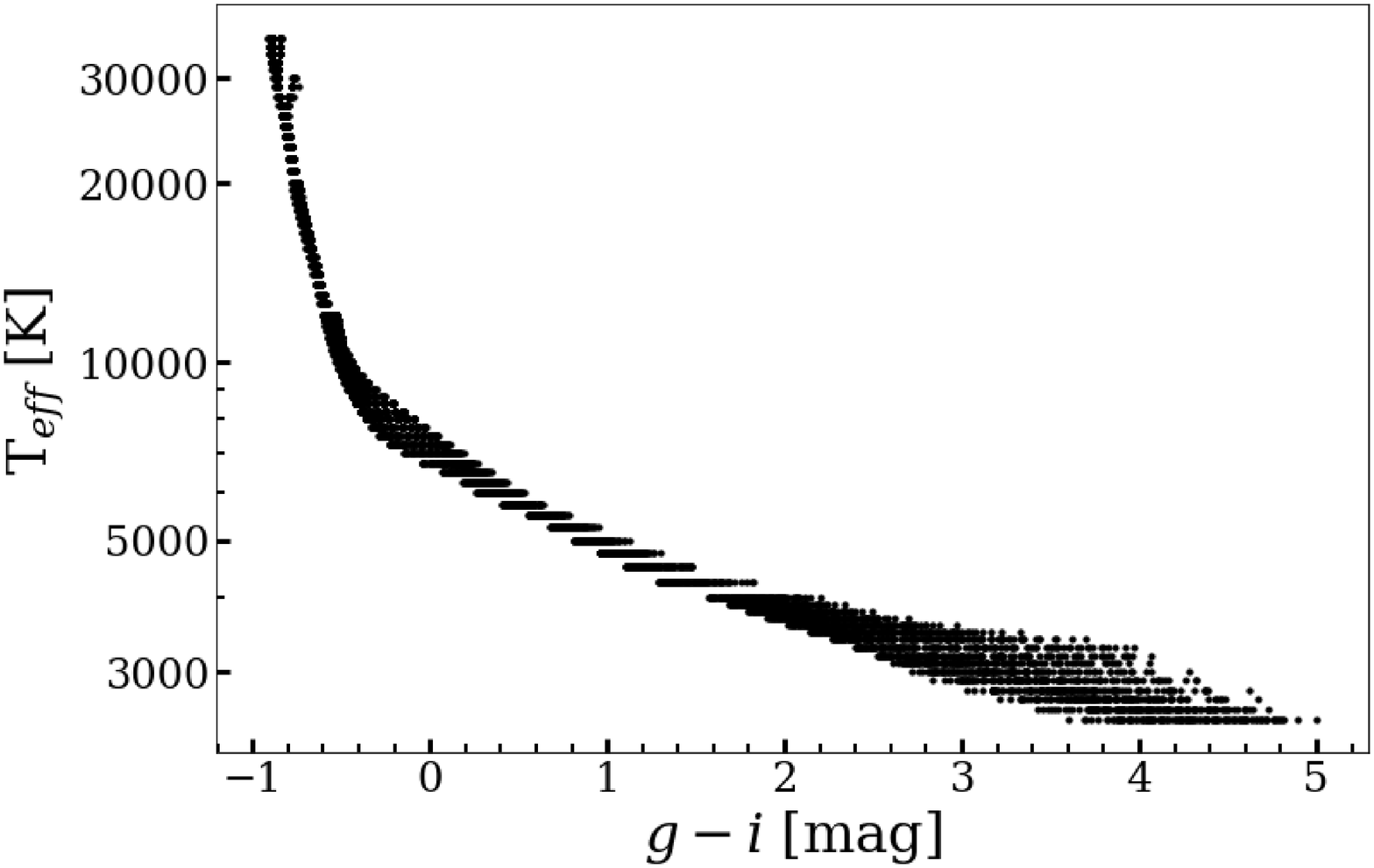}
 \caption{Estimated $g-i$ colours of each theoretical model and their corresponding effective temperature T$_{\rm eff}$.}
 \label{fig:gi_teff}
\end{figure}

To estimate the prior T$_{\rm eff}$, we approximate a function that describes the relationship between the colour ($g - i$) and T$_{\rm eff}$ of the theoretical models. As this is sampled from a discrete grid, a 1-dimensional interpolation function is used to match our observed colour to an effective temperature. The $g - i$ magnitudes are derived from the MaStar spectra for each observation and are provided to the interpolation function which returns the corresponding T$_{\rm eff}$, according to the theoretical models (see Figure~\ref{fig:gi_teff}). This estimate is then used to determine a prior range, whereby an estimate of T$_{\rm eff}<12000\;$K has a range of $\pm1000\;$K and an estimate of T$_{\rm eff}>12000\;$K has a range of $\pm2000\;$K. The prior range is larger for high temperatures as estimates are less accurate at such temperatures and the theoretical grid becomes sparser. 

\subsection{Procedure}
By minimising the $\chi^2$ value between models that fall within our prior range and selected spectra, the FSPs can be obtained. Using this method, the FSPs are derived from individual best-fits and not a combination of models. 

The model fit is performed with pPXF \citep{cappellari04, ppxf17}. Typically used for fitting stellar and gas kinematics in combination with stellar populations of galaxies, here we use this method to fit single stellar templates individually. By using pPXF, one can parameterise the line of sight velocity distribution (LOSVD) of the target and hence account for small velocity offsets in the observed spectrum. The parameterisation of the LOSVD is performed in pPXF using Gauss-Hermite functions (see Section 3.2 in \citet{ppxf17}). Furthermore, inaccuracies in spectral calibration or in reddening effects by dust are accounted for by fitting the observed spectrum with multiplicative polynomials. 

\begin{figure*}
 \includegraphics[width=0.8\paperwidth]{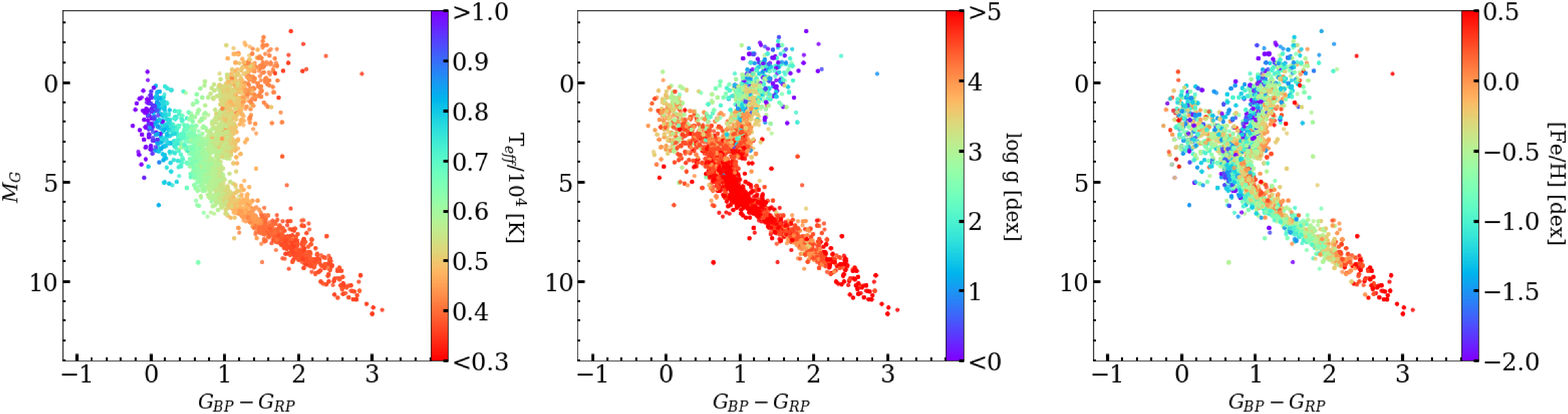}
 \caption{Extinction-corrected MaStar targets (MPL7) shown in a CMD, colour-coded by the recovered FSPs from the discrete $\chi^2$ Method. On the x-axis we use the Gaia colour-index of G$_{BP}-$G$_{RP}$ which represents the blue and red pass bands of the instrument and the y-axis is the Gaia absolute magnitude. The left-hand panel is colour-coded by T$_{\rm eff}$, the middle panel by log g and the right-hand panel by [Fe/H].}
 \label{fig:CMD_results_mpl7}
\end{figure*}

\begin{figure*}
 \includegraphics[width=0.8\paperwidth]{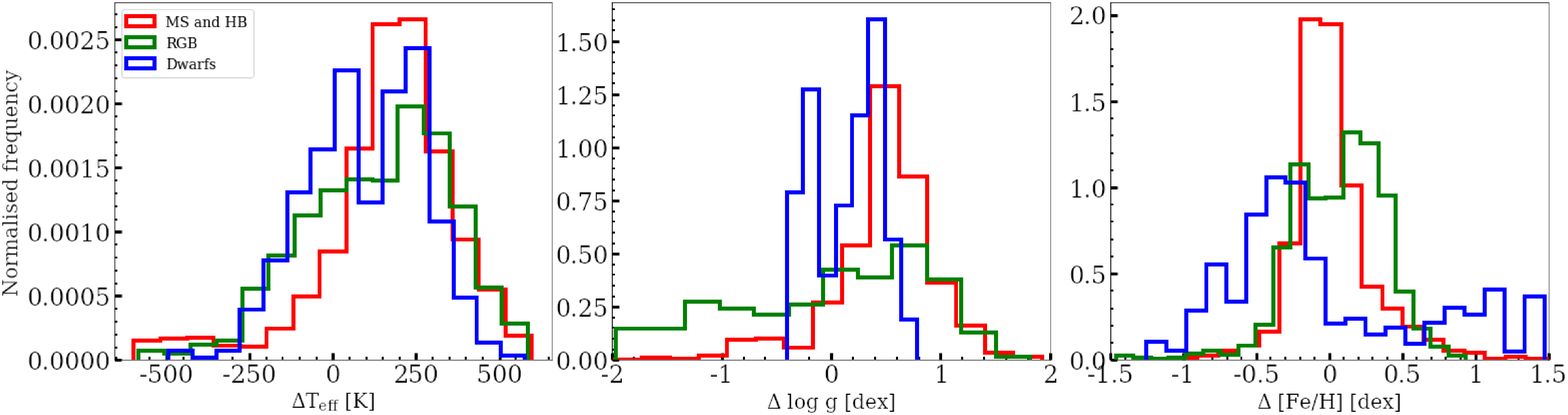}
 \caption{Distribution of the difference between \citet{chen2020} parameters and those from this study using the discrete $\chi^2$ method (this method parameters minus \citet{chen2020}). The comparison is split in to three stellar phases based on our parameters: Dwarfs (T$_{\rm eff} < 5000$ \& log g $\geq4.5$), RGB (T$_{\rm eff} < 6000$ \& log g $< 4.0$) and MS/HB (all remaining spectra).}
 \label{fig:mpl7_comp_yp-lh}
\end{figure*}

\subsection{Results}
\label{sec:chi2_results}
One assessment of the quality of fit of our models can be made by considering the distribution of reduced $\chi^{2}$ values for all observations, which we find the median value to be 8.0 with this method. Despite using a relatively coarse grid of model spectra, we are able to obtain good fits as testified by this low value of the median $\chi^{2}$. The quality of the fits to the data provides some confidence in the recovered parameters. However, by taking the best fitting model, degeneracies in are not mapped.

Using the Gaia photometry-matched values for the MaStar targets we can colour-code CMD plots with the recovered results. Doing so, mislabeled spectra can be identified through our understanding of stellar evolution. The three panels of Figure \ref{fig:CMD_results_mpl7} show the CMD plots for T$_{\rm eff}$, log g and [Fe/H], respectively. On the x-axis we use the Gaia colour-index of $G_{BP}-G_{RP}$ which represents the blue and red pass bands of the instrument and the y-axis is the Gaia absolute magnitude, $M_{G}$. We see good agreement for T$_{\rm eff}$ with bluer colours moving to hotter temperatures. For log g, we expect cool dwarfs to show high values and for this to decrease for decreasing $M_{G}$ due to mass loss and radial expansion towards the giant phase. This is confirmed with the recovered parameters. Furthermore, blue horizontal branch stars may cross the main sequence in the CMD, but the log g values here are different to their main sequence counterparts. This mixing of log g values in this area is expected and the area is populated by stars undergoing rapid stellar evolution. We show how these can be disentangled by plotting parameters in the theoretical Hertzsprung-Russell diagram (HRD) in figure 3 of \citetalias{maraston20}. Finally, metal poor stars are observed to show an ultraviolet excess, due to line blanketing effects, which will increase their T$_{\rm eff}$. We see from the right-hand plot of Figure \ref{fig:CMD_results_mpl7} that as stars move to the bluer part of the CMD, their metallicity generally decreases, with metal-rich targets typically found along the right side of the parameter space, as expected.

For this method applied to DR15 we obtain parameters in the range of: $3400 \leq $ T$_{\rm eff} \leq 25,000\;$K, $0 \leq $ log g $ \leq 5.5\;$dex and $-2.5 \leq [Fe/H] \leq 1.0\;$dex. As this method uses the minimum $\chi^{2}$ value and does not interpolate between model spectra, the errors for each parameter are equal to half the model grid spacing, which varies depending on the location in the grid. The error in T$_{\rm eff}$ is $\pm 50\;$K for T$_{\rm eff} <4250K$, $\pm 125\;$K for $4250<$ T$_{\rm eff}<12000\;$K and $\pm 250\;$K for T$_{\rm eff} \geq 12000\;$K. The error in log g is $\pm 0.25\;$dex for all temperatures. For T$_{\rm eff} < 4250\;$K, the error in [Fe/H] is $\pm 0.25\;$dex for $[Fe/{\rm H}]\leq -1$ and $\pm 0.25\;$dex for $[Fe/{\rm H}] > -1$. For T$_{\rm eff} \geq 4250\;$K, the error in [Fe/H] is $\pm 0.125\;$dex.

We consider the standard error in the parameters for repeat observations of the same star. Doing so, we find mean standard errors for T$_{\rm eff}$, log g and [Fe/H] to be 44 K, 0.07 dex and 0.03 dex, respectively.

\subsubsection{Comparison with \citet{chen2020}}
\label{sec:mpl7-chen-comp}
\citet{chen2020} derive stellar parameters for MPL7 using a semi-empirical approach. They first fit MILES empirical spectra \citep{miles06} with Kurucz model spectra to derive the MILES parameters. They then use the MILES spectra and derived parameters with the ULySS full-spectral fitting package \citep{ulyss09} to fit MaStar spectra based on combinations of the best fitting MILES spectra and their own parameters. For this reason, in \citetalias{maraston20} we refer to this set as 'Empirical' (E-MaStar). Similar to what is presented here, a $\chi^2$ statistic is used to assess quality of fit between the observations and models. However, a key difference is that they use interpolated composite stellar models rather than the single model fit used here. An advantage of the method presented by \citet{chen2020} is that by interpolating between model spectra they are not restricted to their model grid. Nonetheless, they are restricted in overall parameter range by the MILES library.

\begin{table}
\caption{Median difference values between this work and that of \citet{chen2020} for Dwarf, RGB and MS/HB stars.} 
\centering 
\begin{tabular}{c c c} 
\hline\hline 
FSP & Phase & Median difference \\ [0.5ex]
\hline 
T$_{\rm eff}$ (K) & MS/HB & 191 \\ 
 & RGB & 157 \\
 & Dwarf & 107 \\
Log g (dex) & MS/HB & 0.51 \\
 & RGB & 0.14 \\
 & Dwarf & 0.21 \\
$\rm[Fe/H]$ (dex) & MS/HB & -0.02 \\
 & RGB & 0.10 \\
 & Dwarf & -0.24\\[1ex] 
\hline 
\end{tabular}
\label{table:chen-comp-stats} 
\end{table}

In Figure \ref{fig:mpl7_comp_yp-lh} we show a comparison between the atmospheric parameters from our method to those by \citet{chen2020} (our parameters subtract Chen's) which we split in to three stellar phases as defined in Section \ref{sec:lit-comp}. We perform the comparison within the narrower parameter range provided by Chen et al., namely $3376 \leq $ T$_{\rm eff} \leq 19213\;$K. The median offset for each stellar phase and FSP is shown in Table \ref{table:chen-comp-stats}. Firstly, the offset in T$_{\rm eff}$ is shown to have a dependency on stellar type, with largest median difference for MS and HB stars. This stems from the larger uncertainty for hotter stars in the MS. In general, we tend to estimate a higher T$_{\rm eff}$ than Chen. The MS/HB stars show an offset of $0.37$ dex higher than the RGB for log g. Although the offset in log g for the RGB is just 0.14 dex, there is a population of stars where we find lower values than Chen, which reduces the median difference. For [Fe/H] the systematic difference is small for MS/HB and the RGB. We tend to find lower metallicities for dwarf stars with a population where we predict them to be more metal rich by up to 1 dex.

In \citetalias{maraston20} we show that the testing of stellar population models as a function of stellar parameters find consistent results for E-MaStar and Th-MaStar, but favours Th-MaStar for age determination (see details in \citetalias{maraston20}). By performing such tests, we are able to assess the scale of the parameters simultaneously.

We also calculate the standard error for repeat observations as done previously for the parameters we present. For Chen's results, they find a mean standard error of 8 K, 0.02 dex and 0.01 dex for T$_{\rm eff}$, log g and [Fe/H]. As the number of spectra with derived FSPs differ between the two sets, we repeat this measurement for our results using the same sample of spectra. For T$_{\rm eff}$, log g and [Fe/H] we find mean standard errors of 32K, 0.06 dex and 0.03 dex, respectively. In terms of T$_{\rm eff}$, Chen's results are four times more precise than what we find. However, compared the the average temperature of stars in the catalogue, both errors are not significant. The same is true for the errors in log g and [Fe/H].

\section{Comparison of the Discrete \texorpdfstring{$\chi^2$}{k} Method and MCMC Method}
\label{sec:comparison_methods}
Here we present a comparison of the discrete $\chi^2$ and MCMC methods using the MPL7 catalogue of spectra. We assess how well each method can recover the FSPs and a selection of full-spectral fits. For this comparison we also maintain the same procedure for the MCMC Method in terms of the priors, interpolation and algorithm as has been previously described. To clarify, the main difference that the MCMC Method offers is the interpolation of model spectra off of the MARCS and BOSZ-ATLAS9 parameter grid and use of MCMC to probe the posterior distribution of each FSP. One may expect a difference in the recovered parameters due to asymmetric posterior distributions in the MCMC Method and degeneracies between FSPs that will affect the position of the median in the posterior.

\begin{figure*}
 \includegraphics[width=0.8\paperwidth]{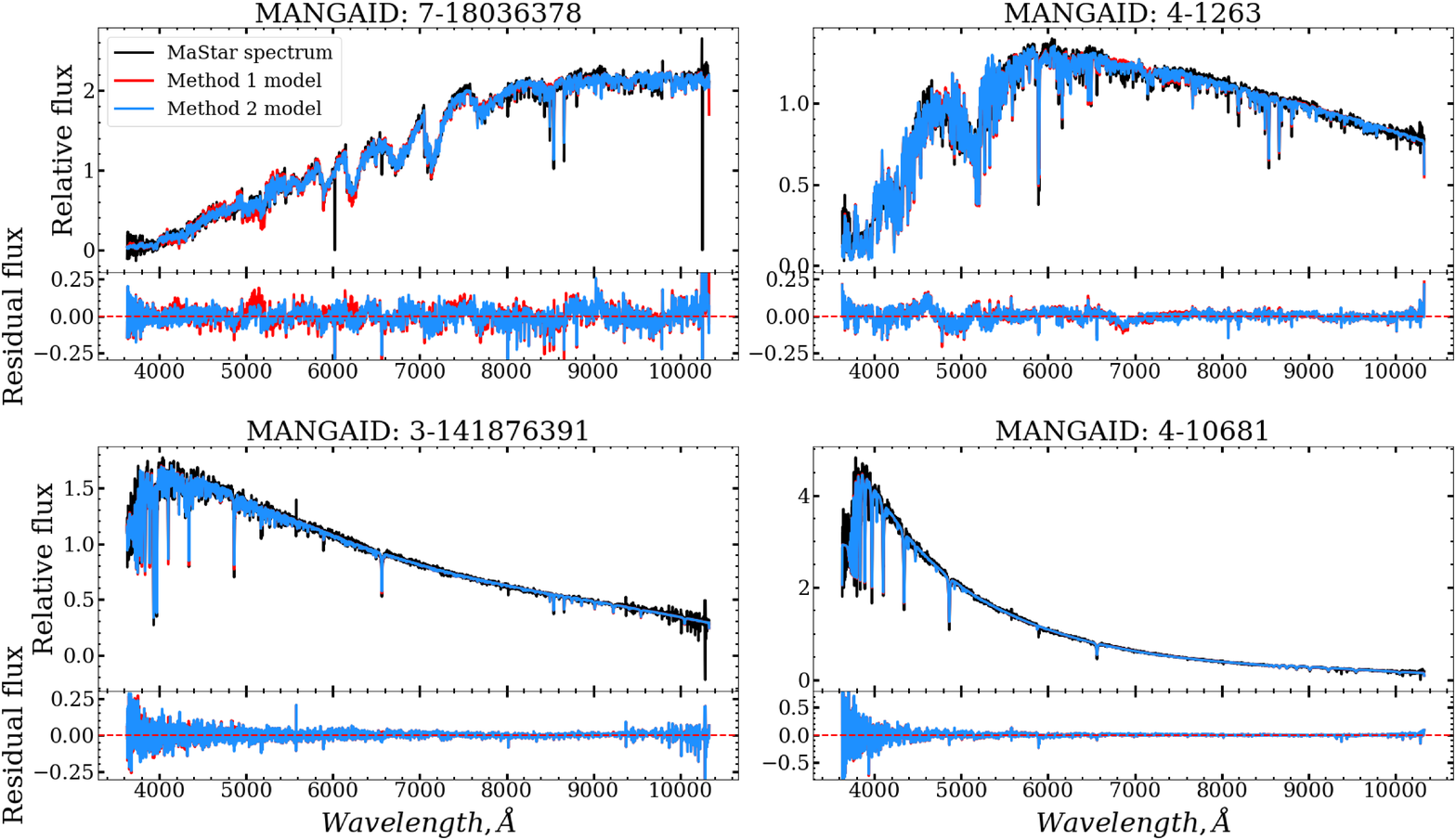}
 \caption{Full-spectral fits of a range of spectral types from the MPL7 MaStar library. We show the fits using the discrete $\chi^2$ Method (red) and the MCMC Method (blue) in the upper panel of each section. The lower panel of each section shows the residual flux (data - model) for each spectral fit.} 
 \label{fig:method_comp_fits}
\end{figure*}

Firstly, we consider stellar fits of different spectral types using both methods. We make an assessment of the recovered parameters and quality of the full-spectral fit. As shown in Figure \ref{fig:method_comp_fits}, the full spectral fit using the MCMC Method is able to recover comparable fits to the discrete $\chi^2$ Method. In Table \ref{table:method_comp} we compare the parameters recovered from these spectral fits. The parameters of the MCMC Method shown here are similar to those of the discrete $\chi^2$ Method, with the most significant discrepancies in [Fe/H]. Of the 8646 spectra in MPL7, we find 4614 that have a $\chi^2 < 30$ for both methods and with corresponding literature values from APOGEE, SEGUE and LAMOST. For the discrete $\chi^2$ Method we recover median differences for T$_{\rm eff}$, log g and [Fe/H] of $-117\;$K, $-0.38\;$dex and $-0.04\;$dex. Using the MCMC Method we find a closer agreement with smaller systematic offsets for all paremeters with median differences of $-65\;$K, $-0.11\;$dex and $0.01\;$dex.

\begin{figure*}
 \includegraphics[width=0.8\paperwidth]{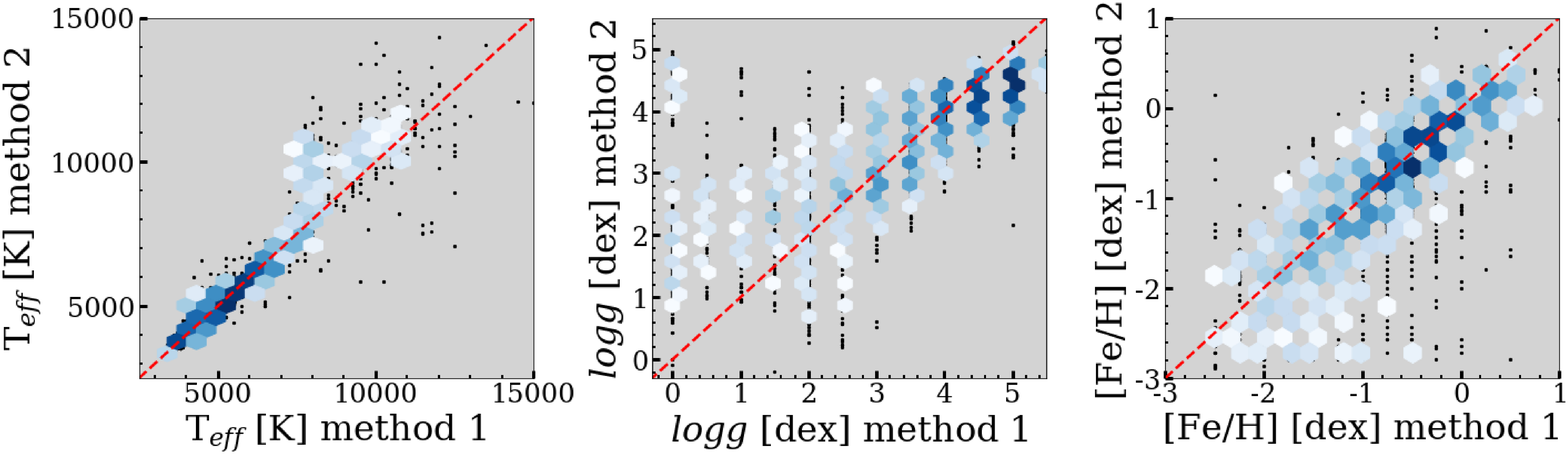}
 \caption{Comparison of FSPs derived from the MaStar MPL7 data release using the discrete $\chi^2$ and MCMC methods. The density of points is shown by the hexagons, where white is a minimum of five data points and dark blue shows higher density. Single black points are low density that aren't included in the minimum bin count of five. The red dashed line is a one-to-one marker that indicates perfect correlation.}
 \label{fig:method_comp_m1-m2}
\end{figure*}

In Figure \ref{fig:method_comp_m1-m2} we show a comparison of the MPL7 FSPs using the discrete $\chi^2$ and MCMC methods. These density plots for the FSPs show the one-to-one relation between the two methods, with denser regions represented by darker blue hexagons. For T$_{\rm eff}$, the MCMC Method estimates slightly hotter values at around 8000K. Standard deviations of $279\;$K and $1249\;$K are found for T$_{\rm eff} < 10,000\;$K and T$_{\rm eff} \geq 10,000\;$K. This difference at high temperatures may be due to the wider priors in T$_{\rm eff}$ used in the MCMC Method. The offset in T$_{\rm eff}$ is small at just $67\;$K. Log g shows the largest discrepancy of all the FSPs, there is a considerable scatter of 0.6 dex and a median offset of 0.4 dex, which is more prominent at lower values. The precision of log g in the model grid is 0.5 dex, which could explain this larger scatter in values. Furthermore, asymmetric posterior distributions at the edge of the model grid lead to a discrepancy in parameter determination. Finally, the metallicity shows some scatter with a standard deviation of $0.4\;$dex and a small systematic median offset of $0.1\;$dex.

\begin{table}
\caption{Comparison of the FSPs recovered from the full-spectral fits of Figure \ref{fig:method_comp_fits}, using the discrete $\chi^2$ and MCMC methods.} 
\centering 
\begin{tabular}{c c c c} 
\hline\hline 
ID & Parameter & Discrete Method & MCMC Method \\ [0.5ex]
\hline 
7-18036378 & T$_{\rm eff}$ (K) & 3400 & 3569 \\ 
 & Log g (dex) & 0.0 & 0.6 \\
 & [Fe/H] (dex) & -1.5 & -0.3 \\
4-1263 & T$_{\rm eff}$ (K) & 4500 & 4536 \\
 & Log g (dex) & 5.0 & 4.7 \\
 & [Fe/H] (dex) & -0.3 & -0.1 \\
3-141876391 & T$_{\rm eff}$ (K) & 6500 & 6281 \\ 
 & Log g (dex) & 4.5 & 4.1 \\
 & [Fe/H] (dex) & -0.5 & -0.6 \\
4-10681 & T$_{\rm eff}$ (K) & 16000 & 15360 \\
 & Log g (dex) & 4.0 & 4.2 \\
 & [Fe/H] (dex) & -0.8 & 0.2 \\[1ex] 
\hline 
\end{tabular}
\label{table:method_comp} 
\end{table}

\section{Additional Parameter Considerations}
\label{sec:additional_params}

\subsection{Microturbulence}
The relationship between atmospheric stellar parameters (T$_{\rm eff}$, log g) and microturbulence ($\xi$) is complex and varies across the H-R diagram \citep{montalban07, aspcap15}. Throughout our analysis we have used a fixed $\xi = 2$ kms$^{-1}$ as this is what's available for the BOSZ models. By fixing $\xi$ we may expect some systematic effects in the results for giants and dwarfs due to the broadening of absorption features. Here we briefly explore the effects of using models with alternative values of $\xi$ when estimating the parameters of dwarf and giant stars. We focus on dwarf and giants as we fit these with MARCS models for which templates at various values of $\xi$ are available. We plan to include $\xi$ as a free parameter in future iterations of the pipeline.

\subsubsection{Giants}
For this experiment we use MARCS models with spherical model geometry, $2500 \leq $ T$_{\rm eff} \leq 8000\;$K, $-0.5 \leq$ log g $\leq 3.5\;$dex and $\xi =$ 2 or 5 kms$^{-1}$. This specific parameter grid reflects what is available at the MARCS website. Initially all spectra with log g $< 2.5\;$dex are selected, we then randomly select ten percent of these to create a representative sample for our analysis. The MCMC method is then applied using the model grids with their respective values of $\xi$ independently.

\begin{figure*}
 \includegraphics[width=0.7\paperwidth]{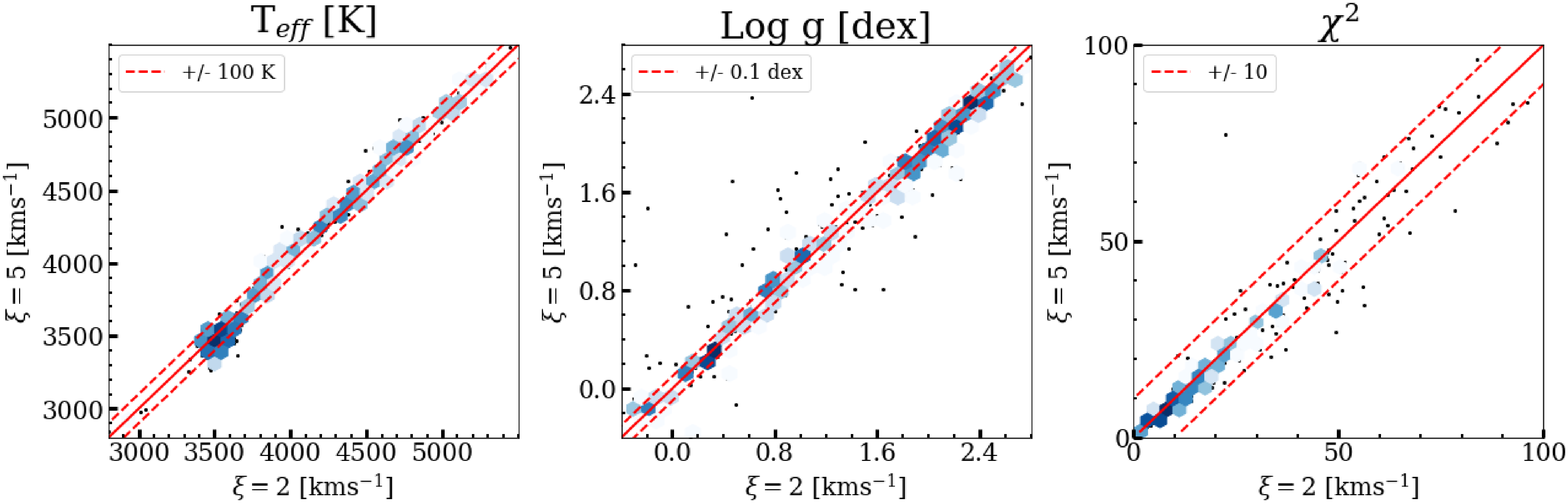}
 \caption{Comparison of T$_{\rm eff}$, log g and $\chi^2$ for giant stars, using the MCMC method and models with a microturbulence of $5$ and $2$ kms$^{-1}$ (our fiducial value). Denser regions are represented by darker blue hexagons and individual spectra by single black points. The red solid line is a one-to-one marker that indicates perfect correlation.}
 \label{fig:microturbulence_giants}
\end{figure*}

In Figure \ref{fig:microturbulence_giants} we show a comparison of the results when using models with $\xi =$ 2 or 5 kms$^{-1}$. For T$_{\rm eff}$ we see some scatter at around $3500\;$K and an offset smaller than 100K towards hotter temperatures when $\xi =$ 5 kms$^{-1}$. Log g shows more scatter and a preference for lower gravity when using $\xi =$ 5 kms$^{-1}$. In terms of the model fit, $\chi^{2}$ values are generally better when using $\xi = 2$ kms$^{-1}$. A median offset of 0.2 is found here.

\begin{figure*}
 \includegraphics[width=0.7\paperwidth]{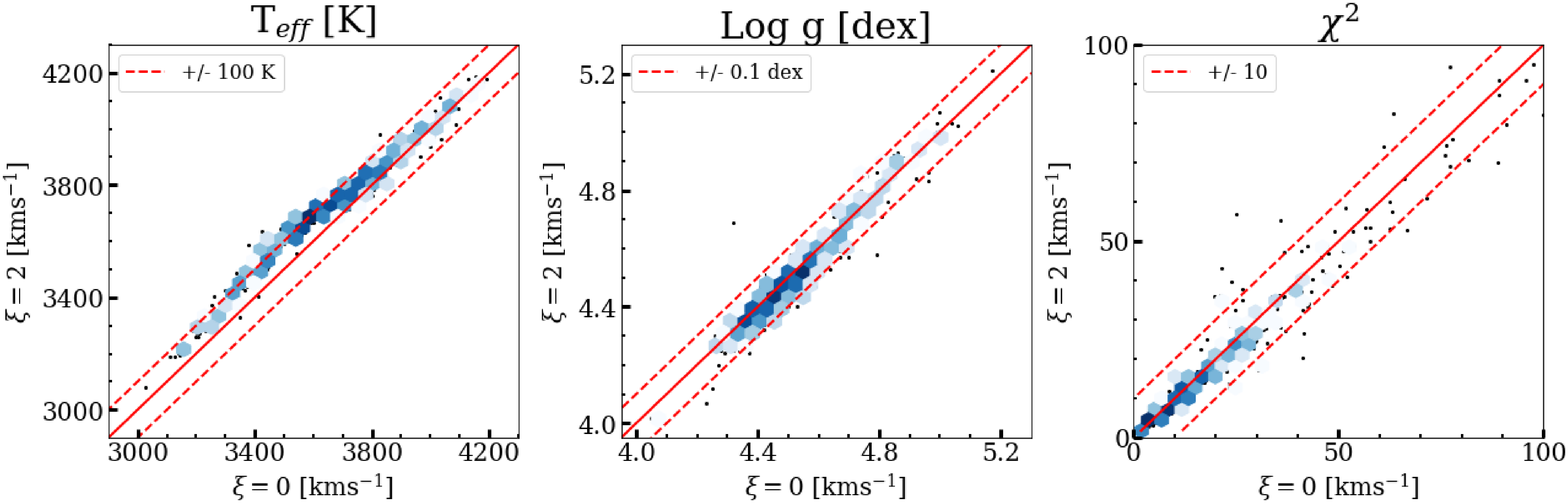}
 \caption{Comparison of T$_{\rm eff}$, log g and $\chi^2$ for dwarfs stars, using the MCMC method and models with a microturbulence of $0$ and $2$ kms$^{-1}$. Denser regions are represented by darker blue hexagons and individual spectra by single black points. The red solid line is a one-to-one marker that indicates perfect correlation.}
 \label{fig:microturbulence_dwarfs}
\end{figure*}

\subsubsection{Dwarfs}
For the analysis of dwarfs we use MARCS models again with plane-parallel model geometry, $2500 \leq $ T$_{\rm eff} \leq 3900\;$K, $3 \leq$ log g $\leq 5\;$dex and $\xi =$ 0 or 2 kms$^{-1}$. The spectra are selected using T$_{\rm eff}$ $\leq 3900\;$K and log g $\geq 4\;$dex and ten percent are used in the analysis.

In Figure \ref{fig:microturbulence_dwarfs} we show a comparison of the results when using models with $\xi =$ 0 or 2 kms$^{-1}$. Regarding T$_{\rm eff}$ we see a systematic offset of approximately $+100\;$K towards the models with $\xi = 2$ kms$^{-1}$ at temperatures below $3800\;$K. For log g there is a slight preference to higher gravity when $\xi = 0$ kms$^{-1}$, but the difference is within $\pm0.1\;$dex. Higher gravity values are in better agreement. Again, the $\chi^{2}$ values are generally smaller for $\xi = 2$ kms$^{-1}$, with a median offset of 0.7.

\subsection{Rotational Velocity}
\begin{figure*}
 \includegraphics[width=0.8\paperwidth]{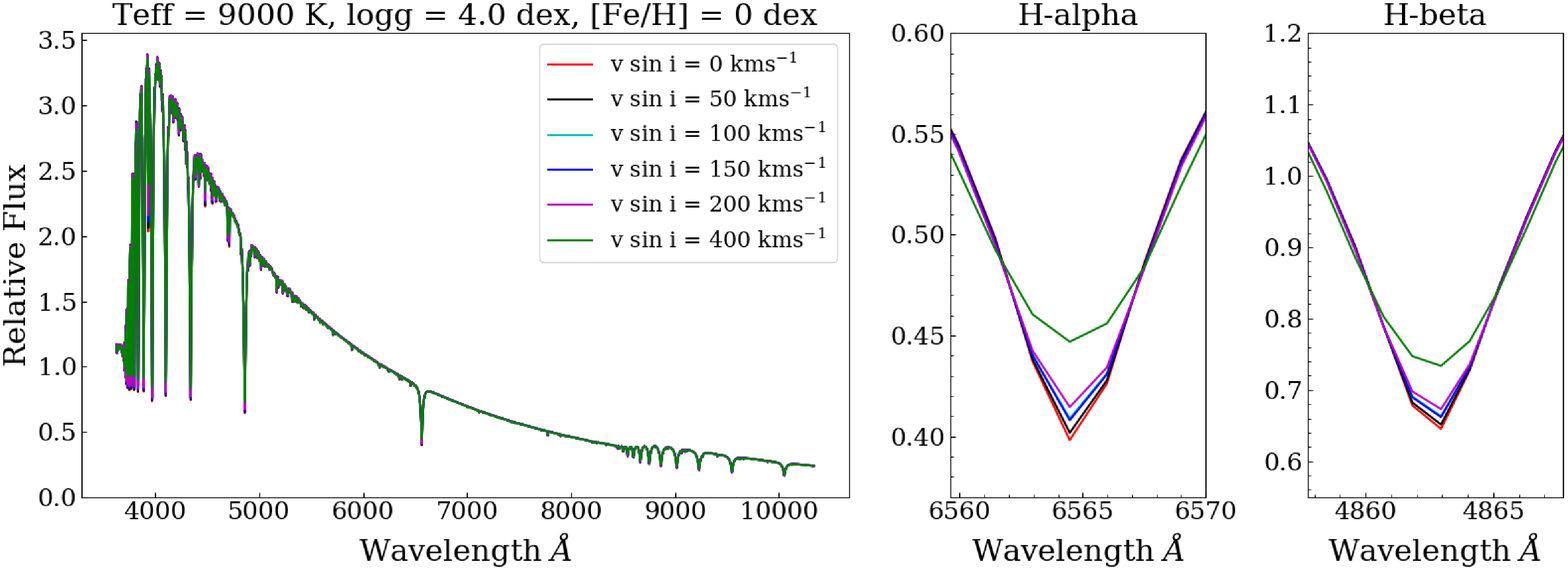}
 \caption{BOSZ synthetic spectra with T$_{\rm eff} = 9000\;$K, log g $= 4\;$dex and [Fe/H] $= 0\;$dex, convolved with rotation profiles of $0 - 400$ kms$^{-1}$. The middle and right panels shows the H-alpha and H-beta absorption lines in detail.}
 \label{fig:vsini_logg_model}
\end{figure*}

\begin{figure*}
 \includegraphics[width=0.8\paperwidth]{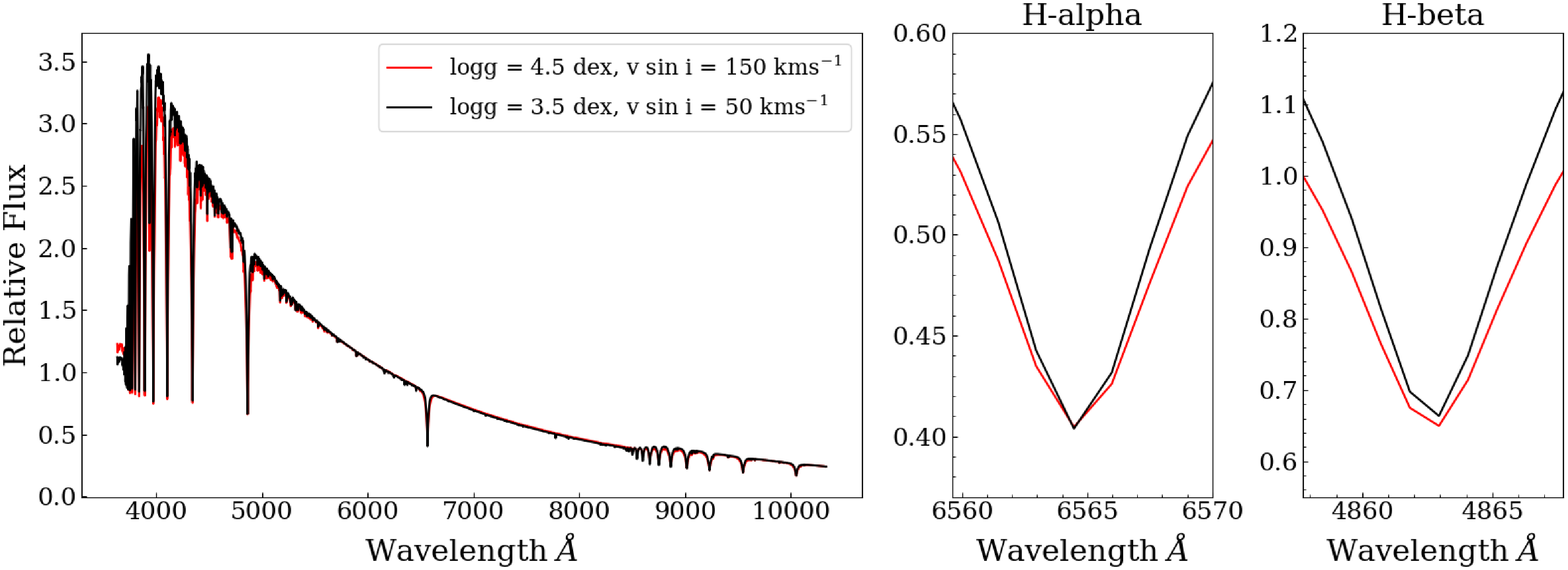}
 \caption{Two BOSZ synthetic spectra with T$_{\rm eff} = 9000\;$K, [Fe/H] $= 0\;$dex. The red spectrum represents log g $= 4.5\;$dex, \textit{v} sin \textit{i} $= 150$ kms$^{-1}$ and the black shows log g $= 3.5\;$dex, \textit{v} sin \textit{i} $= 50$ kms$^{-1}$. The middle and right panels shows the H-alpha and H-beta absorption lines in detail.}
 \label{fig:vsini_logg_variation}
\end{figure*}

The rotational velocity of stars causes a broadening of absorption features and reduces their depth; this may lead to a degeneracy between rotation and log g. To see what effect this would have on the synthetic spectra at MaStar resolution we convolve the BOSZ models with a rotation profile \citep{gray08} for six values of \textit{v} sin \textit{i}: $0, 50, 100, 150, 200, 400$ kms$^{-1}$. In Figure \ref{fig:vsini_logg_model} we show that up to a \textit{v} sin \textit{i} $= 200$ kms $^{-1}$, only the central pixel of absorption lines is affected, and the effect is negligble. In the most extreme case, \textit{v} sin \textit{i} $= 400$ kms$^{-1}$, absorption lines become noticeably shallower and broader. In terms of degeneracy with log g, we show in Figure \ref{fig:vsini_logg_variation} a model with T$_{\rm eff}$ $= 9000\;$K, log g $= 3.5\;$dex, [Fe/H] $= 0\;$ dex and \textit{v} sin \textit{i} $= 50$ kms$^{-1}$ will have the same absorption line depth as a model with log g $= 4.5\;$dex and \textit{v} sin \textit{i} $= 150$ kms$^{-1}$. However, the wings are significantly different which indicates broadening from rotation, not the change in gravity. 

\begin{figure*}
 \includegraphics[width=\columnwidth]{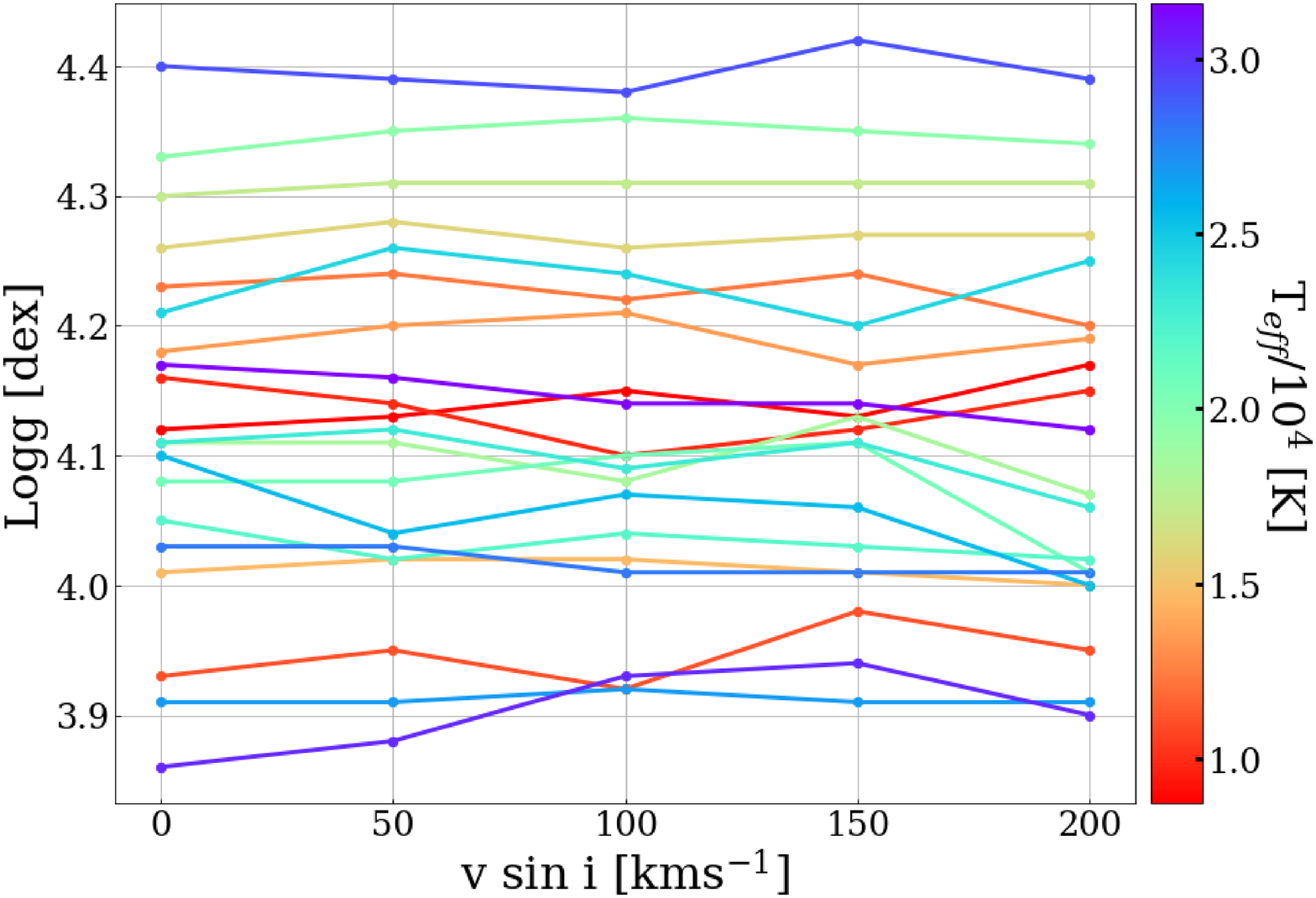}
 \caption{Variation in log g with \textit{v} sin \textit{i} and T$_{\rm eff}$ for a sample of twenty MaStar spectra.}
 \label{fig:vsini_logg}
\end{figure*}

Furthermore, we investigate this effect on twenty MaStar spectra which have been previously analysed by our pipeline with \textit{v} sin \textit{i} $= 0$ kms$^{-1}$. The selected spectra are representative of hot stars ($8700 < $ T$_{\rm eff} < 32000\;$K) as the rotational velocity is typically larger in this temperature range \citep{glebocki05}. We convolve the BOSZ models to obtain spectra with \textit{v} sin \textit{i} = $50 - 200$ kms$^{-1}$. These models are then used independently in our pipeline to determine the fundamental atmospheric parameters. In Figure \ref{fig:vsini_logg} we show the variation in log g with the convolved models and find no systematic effect between the two variables. This result confirms that generally the effect of rotation is not a major concern at MaStar's resolution.

\label{lastpage}

\end{document}